\documentclass[table]{ccjnl}
\usepackage{lipsum,amsmath}
\usepackage{cuted}
\usepackage{bm}
\usepackage{ragged2e}

\title{Reconfigurable Intelligent Surface Aided Integrated Communication and Localization with a Single Access Point}
\author{Xiyu~Wang\inst{1}, Yixuan~Huang\inst{2}, Jie~Yang\inst{3,4,*}, Yu~Han\inst{2}, Shi~Jin\inst{2,4,*} \corinfo{jinshi@seu.edu.cn; yangjie@seu.edu.cn}}
\receiveddate{Mar. 27, 2023}
\reviseddate{Dec. 14, 2022, and Feb. 20, 2023}
\Editor{Tao Jiang}

\address[1]{ZTE Corporation, Shenzhen 518057, China}
\address[2]{National Mobile Communications Research Laboratory, Southeast University, Nanjing 210096, China}
\address[3]{School of Automation, Southeast University, Nanjing 210096, China}
\address[4]{Frontiers Science Center for Mobile Information Communication and Security, Southeast University, Nanjing 210096, China}

\begin{document}

\maketitle

\begin{abstract}
Reconfigurable intelligent surfaces (RISs) not only assist communication but also help the localization of user equipment (UE).
This study focuses on the indoor localization of UE with a single access point (AP) aided by multiple RISs.
First, we propose a two-stage channel estimation scheme where the phase shifts of RIS elements are tuned to obtain multiple channel soundings.
In the first stage, the newtonized orthogonal matching pursuit algorithm extracts the parameters of multiple paths from the received signals.
Then, the LOS path and RIS-reflected paths are identified.
In the second stage, the estimated path gains of RIS-reflected paths with different phase shifts are utilized to determine the angle of arrival (AOA) at the RIS by obtaining the angular pseudo spectrum.
Consequently, by taking the AP and RISs as reference points, the linear least squares estimator can locate UE with the estimated AOAs.
Simulation results show that the proposed algorithm can realize centimeter-level localization accuracy in the discussed scenarios.
Moreover, the higher accuracy of pseudo spectrum, a larger number of channel soundings, and a larger number of reference points can realize higher localization accuracy of UE.
\end{abstract}

\section{Introduction}

Future wireless communication systems are required to not only provide communication with high throughput but also possess the ability of localization and environment sensing \cite{de2021convergent}.
Integrated sensing and communication (ISAC) is becoming a key technique in future systems, which achieves the dual functions of sensing and communication in one system \cite{tan2021integrated}.
With the awareness of the user equipment (UE) location, location-based services such as navigation, augmented reality, and autonomous vehicles can be provided \cite{yang2021model}. 
Moreover, communication capacity and network efficiency can be enhanced with location awareness \cite{yang2021integrated}.
With the development of metasurfaces, the reconfigurable intelligent surface (RIS), which is composed of plenty of sub-wavelength tunable elements, is employed to manipulate radio environments intelligently \cite{tang2020wireless,zhou2022dual}.
Communication and localization performance enhancements are believed to be achieved with the aid of RISs in non-line-of-sight (NLOS) conditions \cite{tang2020mimo,liu2022realization,li2022coverage}, while the hardware cost and energy consumption are relatively low \cite{wang2022reconfigurable}.

\begin{figure*}[t]
\centering
\captionsetup{font=footnotesize}
\begin{subfigure}[]{0.355\linewidth}
\centering
\includegraphics[width=\linewidth]{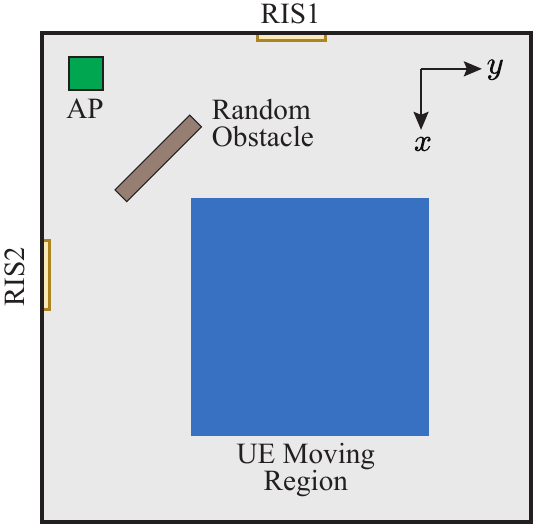}
\caption{Bird View}
\label{fig2a}
\end{subfigure}
\begin{subfigure}[]{0.6\linewidth}
\centering
\includegraphics[width=\linewidth]{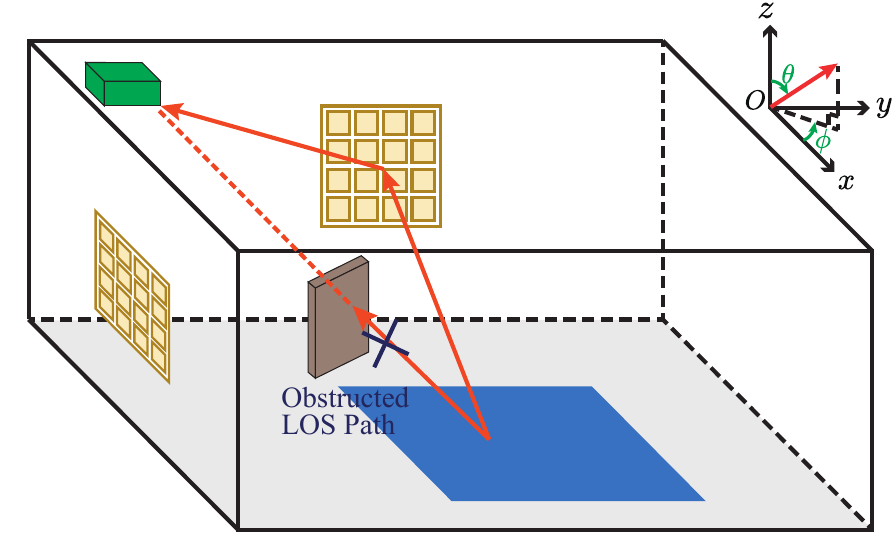}
\caption{3-D View}
\label{fig2b}
\end{subfigure}
\caption{Illustration of the RIS-aided indoor communication systems.}
    \label{fig:model}
\end{figure*}

Recent study has shown an increasing interest in radio-based localization, especially in indoor environments where the GPS may be inoperative \cite{wen2019survey}.
The channel parameters are obtained from channel estimation with the pilot signals and mapped to the unknown UE locations \cite{que2021communication,kordi2020review}.
Then, the linear least squares (LS) estimator can be employed to locate UE, by utilizing the received signal strength (RSS), time of arrival (TOA), time difference of arrival (TDOA), and angle of arrival (AOA) \cite{huang2022joint,dag2018received,guvenc2009survey,jia2018target}.
The cooperation of multiple base stations can obtain rich environmental information from the channel parameters and perform localization with high accuracy \cite{yang2021model}.
However, the RSS and TOA measurements may be inaccurate in indoor environments, owing to the NLOS propagations and synchronization error between UE and AP \cite{yang2021model,katwe2020nlos}.
Considering the directional transmission of uniform planar arrays (UPAs), the estimates of AOA parameters normally deserve high accuracy. 
Therefore, the AOA measurements are preferred for localization in this study.

Distinct from the scenarios considered in \cite{yang2021model,kordi2020review,huang2022joint,dag2018received,guvenc2009survey,jia2018target}, indoor environments usually contain only one access point (AP) \cite{mariakakis2014sail}.
With a single AP, the AOA at AP can only help establish two linear equations, which is unable to determine the 3-D UE location \cite{yang2021model}.
Moreover, the only LOS path may be obstructed by random obstacles.
In literature, the localization and tracking of UE with a single AP can be realized by harnessing the records of motion sensors in smartphones \cite{mariakakis2014sail}.
Additionally, the prior knowledge of the indoor environments (e.g., the distribution of LOS and NLOS regions) can also help locate UE \cite{xie2022simultaneous}.
The rapid development of deep learning techniques enables the channel state information (CSI)-based fingerprinting for localization with a single AP \cite{zhang2019novel}, but a large amount of training data is required for the neural network to learn the environmental information.
To tackle the problems of indoor localization with a single AP, we take advantage of RISs in this study, which requires no extra motion sensors or inconvenient training data collection process, and the only desired environmental information is the locations of the AP and RISs.
Furthermore, the proposed RIS-aided localization scheme coexists with the communication architecture: 
the estimated UE location can help design the phase shifts of RIS elements to enhance the communication capacity and coverage during the communication stage \cite{tang2020mimo}.

In indoor environments, the known locations of RISs make the NLOS conditions with obstructed UE-AP paths transformed into LOS conditions with UE-RIS-AP paths \cite{yildirim2020modeling,basar2019wireless}.
Thus, the RISs also serve as reference points to locate UE.
The Cram\'{e}r-Rao lower bounds (CRLBs) of the estimated UE location and orientation in a single RIS-aided system were derived in \cite{elzanaty2021reconfigurable}, and the CRLB of UE location with obstructed LOS paths was studied in \cite{liu2021reconfigurable}.
However, designing the optimal RIS element phase shifts is a hard work, since the target location should be aware.
Adaptive beamforming of RIS is realized in \cite{he2020adaptive} with the uplink feedback information, leading to extra communication overhead.
RISs are equipped with additional sensors in \cite{shao2022target} to actively sense the locations of possible targets and tune the phase shifts.
In contrast, we randomly choose the phase shifts of RIS elements to avoid large extra feedback information or hardware requirements during the channel estimation stage, while the phase shifts can be manually designed to enhance the communication performance.
Since the propagation distances are limited in indoor environments, the UE-RIS-AP paths with random RIS phase shifts can be separated from environment noise with relatively large transmit power.

Compared with the traditional localization schemes \cite{yang2021model,kordi2020review,huang2022joint,dag2018received,guvenc2009survey,jia2018target,katwe2020nlos,mariakakis2014sail,xie2022simultaneous,zhang2019novel}, the existence of RISs and random phase shifts make the channel estimation difficult, given that the considered RISs in this study can only passively reflect but not actively transmit or receive signals \cite{wei2021channel}.
The radiation field of RIS is divided into the near-field and far-field region via the Rayleigh distance $d_0 = 2D^2/\lambda$, where $D$ is the largest dimension of the RIS, and $\lambda$ is the wavelength.
In the near field of RIS, the distance $d$ between the transmitter/receiver and the RIS should be smaller than $d_0$, and the spherical wave-front over the RIS array is observed.
The problem of channel reconstruction in the near field of RIS has been discussed in \cite{han2022localization,tian2023low}.
Simply, we only consider the far-field conditions in this study, where the plane wave-front is observed, and the channel modelling is relatively easy to follow.
The RISs are deployed on the walls of the indoor environments, and the possible locations of UE are randomly chosen in a given region, as shown in Fig. \ref{fig:model}.
The method of newtonized orthogonal matching pursuit (NOMP) has been proposed to extract the channel parameters of multiple paths in the traditional communication systems \cite{mamandipoor2016newtonized}, and was utilized for the cascaded channel estimation in RIS-aided scenarios \cite{liu2021cascaded}.
Given that the AP and RISs have fixed locations, the RIS-AP channel can be time-invariant.
Therefore, the RIS-reflected paths can be identified from the multiple paths with the knowledge of RIS locations.
On the contrary, the UE-RIS channel may be time-varying owing to the mobility of UE \cite{chen2021low}.
In this study, the pseudo spectrum-based channel estimation scheme proposed in \cite{deepak2020channel} is employed to estimate the AOAs at the RISs by using the path gains of RIS-reflected paths, which can be derived from the NOMP algorithm.
Accordingly, the LS estimator can determine the UE location with the estimated AOAs at the reference points (AP and RISs).

In this study, we mainly focus on the RIS-aided indoor localization with a single AP.
To obtain the channel parameters related to UE location, we propose a two-stage channel estimation scheme, which requires multiple channel soundings with different phase shifts of RIS elements.
In the first stage, the NOMP algorithm proposed in \cite{mamandipoor2016newtonized} is used for the extraction of the multiple paths, then, the LOS path and RIS-reflected paths are identified; in the second stage, the pseudo spectrum-based algorithm in \cite{deepak2020channel} is adopted to estimate the AOAs at the RISs.
Consequently, the estimates of the AOAs at the AP and RISs are employed to locate UE with the LS estimator.

The rest of this paper is organized as follows.
Sec. \ref{sec:system-model} introduces the system model.
Sec. \ref{sec:localization} proposes the two-stage channel estimation scheme and locates UE with the LS estimator.
Sec. \ref{sec:result} presents the simulation results and Sec. \ref{sec:conclusion} concludes this paper.

\begin{figure}
    \centering
    \includegraphics[width=0.4\textwidth]{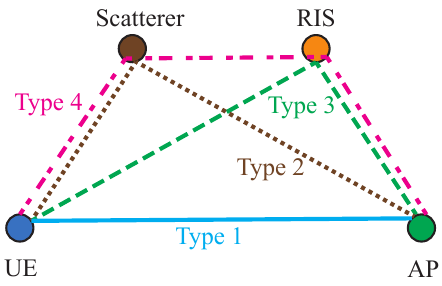}  
    \captionsetup{font=footnotesize}
    \caption{Illustration of the four different types of paths introduced in Sec. \ref{sec:system-model}.}
    \label{fig:paths}
\end{figure}

\section{System Model}
\label{sec:system-model}

We consider a RIS-aided indoor communication system, as shown in Fig. \ref{fig:model}. 
The system works in the 3-D space $\mathbb{R}^{3}=\left\{[x, y, z]^{T}: x, y, z \in \mathbb{R}\right\}$ with the signal wavelength $\lambda$.
The AP is located at $\mathbf{p}=[x_{\rm{p}},y_{\rm{p}},z_{\rm{p}}]^T$, and UE is located at $\mathbf{u}^\circ=[x^\circ,y^\circ,z^\circ]^T$.
The estimated UE location is given as $\mathbf{u}$.
The UE is configured with a single antenna, while the UPA of the AP is composed of $M_{\rm{p}} = M_{\rm{px}}\times M_{\rm{py}}$ antenna elements, with the antenna spacing $d_{\rm{px}} = d_{\rm{py}}=\lambda/2$ \cite{stutzman2012antenna}.
The RISs are deployed on the walls in the indoor environments.
Only the far-field\footnote{note 1} condition is considered in this study, and we aim to estimate $\mathbf{u}^\circ$ by using the received signals at the AP.

We consider four types of paths in this study\footnote{note 2}: UE-AP paths, UE-scatterer-AP paths, UE-RIS-AP paths, and UE-scatterer-RIS-AP paths, as illustrated in Fig. \ref{fig:paths}.
The multipath channel model is given as
\begin{equation}\label{channel}
\mathbf{h} = \mathbf{h}_1 + \mathbf{h}_2 + \mathbf{h}_3 + \mathbf{h}_4 + \mathbf{n},
\end{equation}
where $\mathbf{n}\in\mathbb{C}^{M_{\rm{p}}\times 1}$ is the additive Gaussian noise, and $\mathbf{h}_i\in\mathbb{C}^{M_{\rm{p}}\times 1}$ represents the sub-channel model of the $i$-th type, given as
\begin{equation}\label{h_i}
\mathbf{h}_i = \sum_{l=1}^{L_i} g_{i,l}^\circ \mathbf{a}_{\rm{p}}(\omega_{i,l}^\circ , \psi_{i,l}^\circ ).
\end{equation}
Here, $L_i$ is the number of paths of the $i$-th type; $g_{i,l}^\circ$ is the complex gain of the $(i,l)$-th path, \textit{i.e.}, the $l$-th path of the $i$-th type; $\mathbf{a}_{\rm{p}}(\omega_{i,l}^\circ, \psi_{i,l}^\circ)$ is the steering vector at the AP of the $(i,l)$-th path, given as
\begin{equation}\label{eq:upa-steering}
\begin{aligned}
\mathbf{a}_{\rm{p}}(\omega_{i,l}^\circ, \psi_{i,l}^\circ)& = \mathbf{a}_{\rm{px}}(\omega_{i,l}^\circ)\otimes \mathbf{a}_{\rm{py}}(\psi_{i,l}^\circ),\\
& = \left[1, e^{-j\omega_{i,l}^\circ}, \ldots, e^{-j(M_{\rm{px}} - 1)\omega_{i,l}^\circ}\right]^H \otimes \\
& \ \ \ \ \ \left[1, e^{-j\psi_{i,l}^\circ}, \ldots, e^{-j(M_{\rm{py}} - 1)\psi_{i,l}^\circ}\right]^H,
\end{aligned}
\end{equation}
where $\otimes$ represents the Kronecker product; $j$ is the imaginary unit; $\omega_{i,l}^\circ = 2\pi\frac{d_{\rm{px}}}{\lambda}\cos\phi_{i,l}^{\rm{p}\circ}\sin\theta_{i,l}^{\rm{p}\circ}$ and $\psi_{i,l}^\circ$
$= 2\pi\frac{d_{\rm{py}}}{\lambda}\sin\phi_{i,l}^{\rm{p}\circ}\sin\theta_{i,l}^{\rm{p}\circ}$ are the normalized AOAs at the AP;
$(\phi_{i,l}^{\rm{p}\circ},\theta_{i,l}^{\rm{p}\circ})$ are the azimuth and elevation of AOAs at the AP of the $(i,l)$-th path.
Thus, the steering vector at the AP can be obtained as long as the positions of UE, scatterer, or RIS are given.
Then, we model the complex path gains of different paths.

\begin{itemize}
\item \textbf{Type 1: UE-AP path (LOS path)}: 
The direct path gain between the UE and AP is given as \cite{goldsmith2005wireless}
\begin{equation}
g_{1,l}^\circ = \frac{\lambda \sqrt{G_{\rm{p}}} }{4 \pi d_{\rm{up}}}e^{-\frac{j 2 \pi d_{\rm{up}}}{\lambda}},
\end{equation}
where $G_{\rm{p}}$ is the antenna gain of the receiving antennas at the AP; $d_{\rm{up}}$ represents the distance between the UE and AP.
The number of LOS paths $L_1\in \{0,1\}$, where $L_1=0$ represents the condition that the LOS path is obstructed.

\item \textbf{Type 2: UE-scatterer-AP path (scattered path)}: 
The gain of the $(2,l)$-th scattered path is given as \cite{goldsmith2005wireless}
\begin{equation}
\label{eq:nlos-pathgain-scatterer}
g_{2,l}^\circ = \frac{\lambda \sqrt{G_{\rm{p}} \sigma_{2,l}} }{(4 \pi)^{\frac{3}{2}} d_{2,l}^{\rm{sp}} d_{2,l}^{\rm{su}}}e^{-\frac{j 2 \pi (d_{2,l}^{\rm{sp}} + d_{2,l}^{\rm{su}})}{\lambda}},
\end{equation}
where $d_{2,l}^{\rm{sp}}$ is the distance between the scatterer and AP; $d_{2,l}^{\rm{su}}$ is the distance between the scatterer and UE; $\sigma_{2,l}$ denotes the radar cross section (RCS) of the $(2,l)$-th scatterer.
Considering that the RCSs of scatterers may vary independently from time to time, we adopt the classical Swerling Case II target model in this study \cite{mahafza2017introduction}.

\item \textbf{Type 3: UE-RIS-AP path (RIS-reflected path)}: 
We assume that the RIS is composed of $M_{\rm{r}} = M_{\rm{rx}} \times M_{\rm{ry}}$ tunable elements, with the area of each RIS element $d_{\rm{rx}}\times d_{\rm{ry}}$.
The gain of each RIS element of the $(3,l)$-th path in the far-field region is given as \cite{tang2022path}
\begin{equation}\label{eq:path2-gain-sub-farfield}
g_{3,l}^{\rm{s}\circ}\!=\!\frac{\lambda\!\sqrt{\!d_{\rm{rx}}d_{\rm{ry}} G_{\rm{p}} G_{\rm{r}} F(\varphi_{3,l}^{\rm{rp}\circ}\!,\vartheta_{3,l}^{\rm{rp}\circ})F(\varphi_{3,l}^{\rm{ru}\circ}\!,\vartheta_{3,l}^{\rm{ru}\circ})}}{(4 \pi)^{\frac{3}{2}} d_{3,l}^{\rm{rp}} d_{3,l}^{\rm{ru}}},
\end{equation}
where $G_{\rm{r}}=4\pi d_{\rm{rx}}d_{\rm{ry}} / \lambda^2 $ is the scattering gain of the RIS elements;
$(\varphi_{3,l}^{{\rm{rp}}\circ}, \vartheta_{3,l}^{{\rm{rp}}\circ})$ and $(\varphi_{3,l}^{{\rm{ru}}\circ}, \vartheta_{3,l}^{{\rm{ru}}\circ})$ denote the azimuth and elevation angles pointing from the center of the $(3,l)$-th RIS to AP and UE, and the corresponding distances are given by ${d}_{3,l}^{\rm{rp}}$ and ${d}_{3,l}^{\rm{ru}}$, respectively;
$F(\varphi_{3,l}^{{\rm{rp}}\circ},\vartheta_{3,l}^{{\rm{rp}}\circ})$ and $F(\varphi_{3,l}^{{\rm{ru}}\circ},\vartheta_{3,l}^{{\rm{ru}}\circ}) $ are the normalized power radiation patterns of the RIS element in the directions of reflecting and receiving, with an example as \cite{stutzman2012antenna}
\begin{equation}
\label{eq:pattern}
F(\varphi, \vartheta)\!=\!\left\{\begin{aligned}
&\cos \beta, & &\!\!\beta \in\left[0, \frac{\pi}{2}\right], \\[-2pt]
&0, & &\!\!\beta \in\left(\frac{\pi}{2}, \pi\right],
\end{aligned}\right.
\end{equation}
where $\beta$ denotes the angle between the direction $(\varphi, \vartheta)$ and the normal vector of the RIS array. As illustrated in Fig. 1, $\beta = \arccos (\cos\varphi\sin\vartheta)$ for RIS1, whereas $\beta = \arccos (\sin\varphi\sin\vartheta)$ for RIS2.
Given that the RIS elements are deployed as a UPA, we sum up the path gains of each RIS element by considering the tunable phase shifts and inter-element path length difference, deriving the total path gain of the $(3,l)$-th RIS-reflected path as
\begin{equation}\label{eq:path2-gain-sum-farfield}
g_{3,l}^\circ=g_{3,l}^{\rm{s}\circ}\mathbf{a}_{\rm{r}}^H(\varphi_{3,l}^{\rm{rp}\circ},\vartheta_{3,l}^{\rm{rp}\circ})\boldsymbol{\Omega}_l\mathbf{a}_{\rm{r}}(\varphi_{3,l}^{\rm{ru}\circ},\vartheta_{3,l}^{\rm{ru}\circ}),
\end{equation}
where $\boldsymbol{\Omega}_l=\operatorname{diag}(\xi_{1}^l, \ldots, \xi_{m_{\rm{r}}}^l, \ldots, \xi_{M_{\rm{r}}}^l)\in\mathbb{C}^{M_{\rm{r}}\times M_{\rm{r}}}$ is the phase shifts of RIS elements;
$\xi_{m_{\rm{rx}},m_{\rm{ry}}}^l$ represents the tunable phase shift of the $(m_{\rm{rx}},m_{\rm{ry}})$-th RIS element, \textit{i.e.}, the element located at the $m_{\rm{rx}}$-th row and the $m_{\rm{ry}}$-th column in the RIS array;
$\mathbf{a}_{\rm{r}}(\varphi_{3,l}^{\rm{rp}\circ},\vartheta_{3,l}^{\rm{rp}\circ})\in\mathbb{C}^{{M_{\rm{r}}\times1}}$ and $\mathbf{a}_{\rm{r}}(\varphi_{3,l}^{\rm{ru}\circ},\vartheta_{3,l}^{\rm{ru}\circ})\in\mathbb{C}^{{M_{\rm{r}}\times1}}$ are the steering vectors of the $(3,l)$-th RIS at the directions pointing to the AP and UE, respectively.
The path gain model given by \eqref{eq:path2-gain-sum-farfield} is applicable for all the UE-RIS-AP paths.

\begin{figure*}
    \centering
    \includegraphics[width=0.98\textwidth]{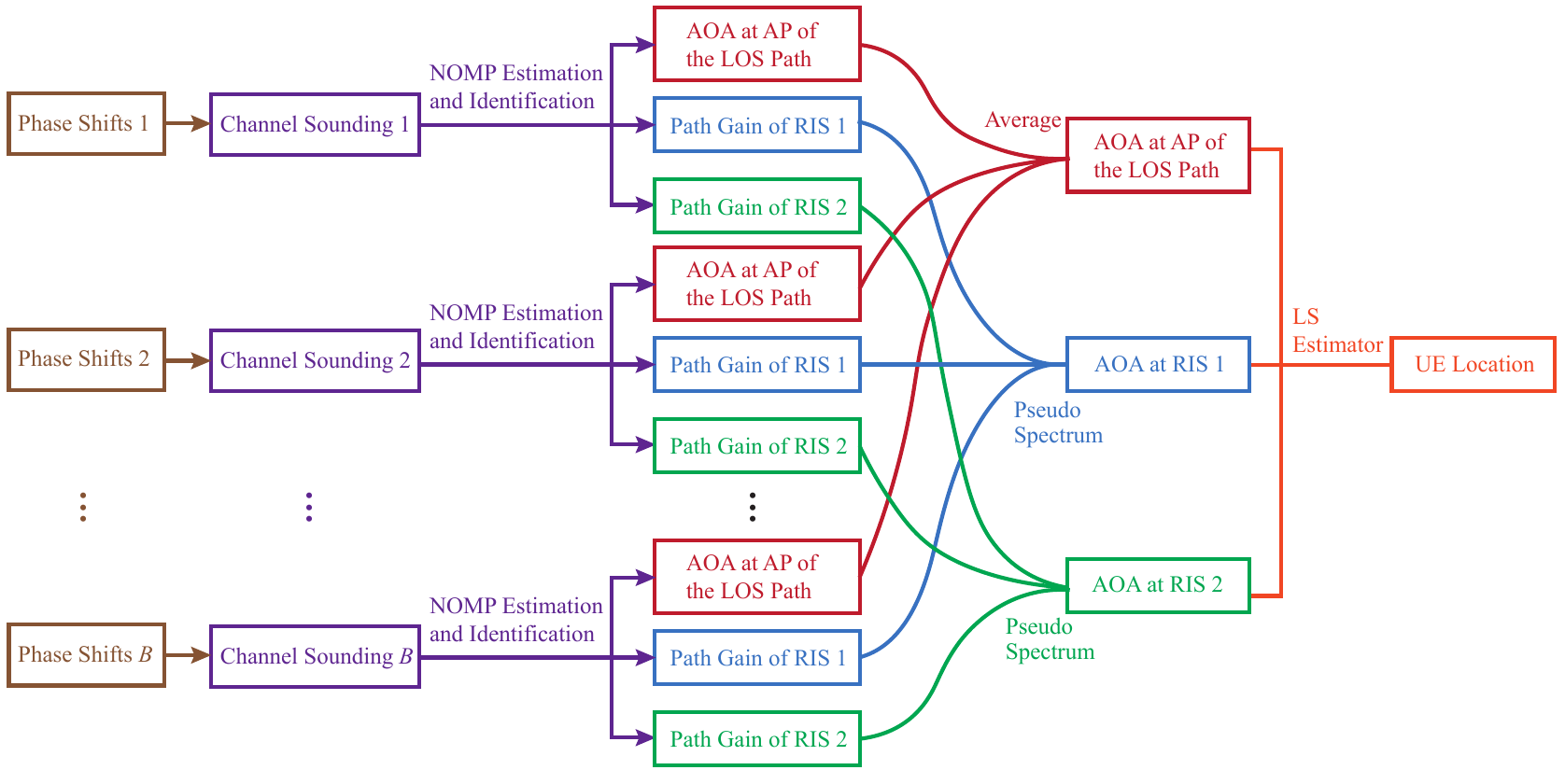} 
    \captionsetup{font=footnotesize}
    \caption{Illustration of the proposed algorithm flow, where ``NOMP Estimation and Identification'', ``Pseudo Spectrum'', and ``LS Estimator'' refer to Sec. \ref{sec-nomp}, Sec. \ref{sec-pseudo}, and Sec. \ref{sec-ls}, respectively.}
    \label{fig:flow}
\end{figure*}

\item \textbf{Type 4: UE-scatterer-RIS-AP path}: 
We divide the $(4,l)$-th path into two parts: in the first part, the signal is transmitted from UE to the scatterer and scattered by the scatterer; in the second part, the signal propagates from the scatterer to the RIS, being reflected by RIS and received by AP.
For the first part, the signal experiences the path loss of free space from the UE to scatterer and that of scattering.
Thus, the path gain of the first part of the $(4,l)$-th path can be given as 
\begin{equation}\label{eq:path3-gain-part}
g_{4,l}^{1\circ} = \frac{\sqrt{\sigma_{4,l}}}{(4 \pi)^{\frac{1}{2}} d_{4,l}^{\rm{su}}}e^{-j2\pi\frac{d_{4,l}^{\rm{su}}}{\lambda}}.
\end{equation}
For the second part, it can be treated as a special kind of path of type 3, where the UE is equivalently located at the $(4,l)$-th scatterer.
Consequently, the path gain of each RIS element of the $(4,l)$-th path is given as
\begin{equation}
g_{4,l}^{\rm{s}\circ}=g_{4,l}^{1\circ}g_{3,l}^{\rm{s}'\circ},
\end{equation}
where $g_{3,l}^{\rm{s}'\circ}$ is in the same form as \eqref{eq:path2-gain-sub-farfield}, but $(\varphi_{4,l}^{\rm{ru}\circ}, \vartheta_{4,l}^{\rm{ru}\circ})$ and $d_{3,l}^{\rm{ru}}$ are replaced by $(\varphi_{4,l}^{\rm{rs}\circ}, \vartheta_{4,l}^{\rm{rs}\circ})$ and $d_{4,l}^{\rm{rs}}$, respectively.
Here, $(\varphi_{4,l}^{\rm{rs}\circ}, \vartheta_{4,l}^{\rm{rs}\circ})$ represent the azimuth and elevation angles pointing from the center of the $(4,l)$-th RIS to scatterer;
$d_{4,l}^{\rm{rs}}$ is the distance between the $(4,l)$-th RIS and scatterer.
As an analogy to \eqref{eq:path2-gain-sum-farfield}, the total path gain of the $(4,l)$-th path is given as
\begin{equation}\label{eq:path3-gain-sum-farfield}
g_{4,l}^\circ=g_{4,l}^{\rm{s}\circ}\mathbf{a}_{\rm{r}}^H(\varphi_{4,l}^{\rm{rp}\circ},\vartheta_{4,l}^{\rm{rp}\circ})\boldsymbol{\Omega}_l\mathbf{a}_{\rm{r}}(\varphi_{4,l}^{\rm{rs}\circ},\vartheta_{4,l}^{\rm{rs}\circ}).
\end{equation}
The path gain model given by \eqref{eq:path3-gain-sum-farfield} is applicable for all the UE-scatterer-RIS-AP paths.

\end{itemize}

To summarize, the multipath channel model has been presented.
Then, channel estimation schemes are employed to estimate the AOAs at the AP and RISs, and the LS estimator is utilized to determine the UE location in the next section.

\section{RIS-aided Indoor Localization}
\label{sec:localization}

In this section, we perform channel estimation and localization in turn, where the localization is based on the known locations of the reference points, \textit{i.e.}, AP and RISs, and the estimated AOAs at them.
To estimate the AOAs, we propose a two-stage channel estimation scheme that requires $B$ ($B\ge2$) channel soundings with different phase shifts.
In the first stage, we employ the NOMP algorithm \cite{mamandipoor2016newtonized} to estimate the channel parameters of multiple paths, which help identify the LOS path and RIS paths in each channel sounding.
Then, the pseudo spectrum-based algorithm in \cite{deepak2020channel} is adopted in the second stage to estimate the AOAs at the RISs.
Finally, the location of UE is determined with the LS estimator.
The algorithm flow chart is depicted in Fig. \ref{fig:flow}.

Here, the phase shifts of RISs are required to be chosen randomly such that the configuration matrix $\boldsymbol{\Omega}_{l,b}$ varies among the $B$ channel soundings.
Notably, we do not require that the phase shifts of all RIS elements are changed between different channel soundings.
Even for the RISs with 1-bit quantization phase shifts, $2^{M_{\rm{r}}}$ different configurations are available to acquire $B$ channel soundings.

\subsection{Estimation and Identification of LOS path and RIS Paths}
\label{sec-nomp}

This subsection (first-stage channel estimation) employs the NOMP algorithm to extract the multiple paths from the complex channel vector given in \eqref{channel}, followed by which the LOS path and RIS paths can be identified.
Given that estimating the AOAs at the RISs requires $B$ channel soundings with different phase shifts, the NOMP algorithm is employed repeatedly for $B$ times.
The NOMP is an iterative algorithm and estimates the parameters of the multiple paths in the descending order of the path power.
Each iteration of the algorithm obtains the path gain and the AOA at the AP of a single path.
Thus, we only present how the algorithm works in a single iteration in this subsection.
We define the maximum number of extracted paths as $L_{\rm{max}}$.
The NOMP algorithm is repeated until the parameters of $L_{\rm{max}}$ paths have been estimated, or the power of the residual channel response is lower than the threshold $E_0$.

Each iteration of the NOMP algorithm includes three steps: detection, single refinement, and cyclic refinement.
In the first step, the NOMP algorithm extracts the path that has the largest power in the residual channel response.
Assuming that $K$ paths have been extracted with the estimated parameters $\mathcal{P}_{b,K}=\left\{\left(g_{b,k}, \omega_{b,k}, \psi_{b,k}\right), k=1, \ldots, K\right\}$, the residual channel response is given as
\begin{equation}\label{residual}
\mathbf{h}_{b,K}^{\rm{r}} = \mathbf{h}_b - \sum_{k=1}^K g_{b,k} \mathbf{a}_{\rm{p}}(\omega_{b,k}, \psi_{b,k}),
\end{equation}
where $\mathbf{h}_b$ is the measured channel response in the $b$-th channel sounding.
The estimation of the $(K+1)$-th path is equivalent to finding the unit-norm steering vector that has the largest projection on $\mathbf{h}_{b,K}^{\rm{r}}$ from a discrete codebook.
The angle corresponding to the steering vector is the estimated AOA at the AP, and the length of the projection is the estimated path gain.
However, the estimates of AOA in the first step only fall in the discrete points in the angle domain, which may result in large estimation errors with a small codebook or large computation overhead with a large codebook.
To overcome this shortcoming, the second step takes a Newton-based refinement stage, allowing the estimates of AOA to fall in the continuous interval $[0,2\pi)$, to degrade the energy of the residual channel response as much as possible.
Followed by this, the third step refines the parameters of all the estimated paths one by one with the same method as the second step.
The second and third steps are repeated for $R_{\rm{s}}$ and $R_{\rm{c}}$ times, respectively.
After $K_0$ ($K_0 \le L_{\rm{max}}$) iterations of the NOMP algorithm, the estimates of the channel parameters of the $b$-th channel sounding are given as $\mathcal{P}_{b,K_0} = \left\{\left(g_{b,k}, \omega_{b,k}, \psi_{b,k}\right), k=1, \ldots, K_0\right\}$.

Then, we identify the channel parameters that correspond to the LOS path and RIS paths.
Since the locations of the AP and RISs are known, we can obtain the true normalized AOA at the AP of the $l$-th RIS path, given as 
\begin{equation}
\begin{aligned}
(\omega_{3,l}^{\circ}, \psi_{3,l}^{\circ}) =  \text{mod}&\left[\left( 2\pi \frac{d_{\rm{px}}}{\lambda} \cos\phi_{3,l}^{\rm{p}\circ} \sin\theta_{3,l}^{\rm{p}\circ}, \right. \right.\\
& \left. \left. 2\pi \frac{d_{\rm{py}}}{\lambda} \sin\phi_{3,l}^{\rm{p}\circ} \sin\theta_{3,l}^{\rm{p}\circ} \right), 2\pi\right].
\end{aligned}
\end{equation}
Thus, we can obtain the path that has the minimum estimation error with $(\omega_{3,l}^{\circ}, \psi_{3,l}^{\circ})$ from $\mathcal{P}_{b,K_0}$.
This path is the estimate of the $l$-th RIS path if the estimation error $T_{l,k}$ is smaller than the threshold $T_0$.
Then, we can obtain the estimated path gain $g_{3,l,b}$.
Otherwise, the $l$-th RIS path in the $b$-th channel sounding is unable to be estimated, which may result from an obstructed RIS path or large measurement noise.
Here, the estimation error $T_{l,k}$ between the $l$-th RIS path and the $k$-th estimated path is given as 
\begin{equation}\label{111}
T_{l,k} =  |\omega_{3,l}^{\circ} - \omega_{b,k}|^2 + |\psi_{3,l}^{\circ} - \psi_{b,k}|^2.
\end{equation}

Consequently, we identify the LOS path after extracting the RIS paths.
Since the power of the LOS path is usually the largest, we regard the first path as the LOS path if it has not been assigned to any RIS paths.
Otherwise, the measured channel response $\mathbf{h}_{b}$ includes no LOS paths.
If the LOS path can be detected, we denote the estimates of the normalized AOA at the AP as $(\omega_{1,1,b}, \psi_{1,1,b})$.
Then, the estimated AOA at the AP of the $b$-th channel sounding can be given by
\begin{equation}\label{222}
\begin{aligned}
&\phi_{1,1,b}^{\rm{p}} = \arctan\frac{d_{\rm{px}}\psi_{1,1,b}}{d_{\rm{py}}\omega_{1,1,b}},\\
&\theta_{1,1,b}^{\rm{p}} = \arcsin \frac{\lambda\omega_{1,1,b}}{2\pi d_{\rm{px}}\cos\phi_{1,1,b}}.
\end{aligned}
\end{equation}
Notably, any position in the UE moving region shown in Fig. \ref{fig:model} ensures that $\phi_{1,1,b}^{\rm{p}\circ} \in [0, \frac{\pi}{2}]$ and $\theta_{1,1,b}^{\rm{p}\circ} \in [0, \pi]$, thus the solution given in \eqref{222} matches the practical environments.
Additionally, we assume that the LOS path can be detected from $B_0$ channel soundings.
Then, the final estimates of the AOA of the LOS path can be calculated by averaging, given as
\begin{equation}\label{average}
\phi_{1,1}^{\rm{p}} = \frac{1}{{B_0}}\sum_{b=1}^{B_0 }\phi_{1,1,b}^{\rm{p}},\  \theta_{1,1}^{\rm{p}} = \frac{1}{{B_0}}\sum_{b=1}^{B_0 }\theta_{1,1,b}^{\rm{p}}.
\end{equation}

To summarize, the NOMP algorithm can estimate the AOA of the LOS path at the AP $(\phi_{1,1}^{\rm{p}}, \theta_{1,1}^{\rm{p}})$ and extract the path gains of the RIS paths $g_{3,l,b}$ from each channel sounding.
In the next subsection, we determine the AOA of the RIS-reflected paths at the RISs based on the estimates $g_{3,l,b}$, where $l=1,2,\ldots,L_3$ and $b=1,2,\ldots,B$.

\subsection{Estimation of the AOAs at RISs}
\label{sec-pseudo}

This subsection (second-stage channel estimation) works on the estimation of the AOAs at the RISs based on the estimates of the NOMP algorithm.
Since the multiple paths in the channel response $\mathbf{h}_b$ are identified by their AOAs at AP in Sec. \ref{sec-nomp}, the paths of type three and type four that correspond to the same RIS can not be separated.
Therefore, the estimated path gain $g_{3,l}$ in Sec. \ref{sec-nomp} consists of the paths that are reflected by the $l$-th RIS (including type three and type four). 
However, the path gains of type four are typically much smaller than that of type three.
Under the system configurations in Sec. \ref{sec:result}, the energy of the signals of type four is around 20 dB lower than that of type three.
Thus, we treat the estimated path gain $g_{3,l}$ as that of the $l$-th UE-RIS-AP path and treat the paths of the fourth type as interference to the system.
In this way, we can estimate the AOAs at the RISs by utilizing the pseudo spectrum-based algorithm proposed in \cite{deepak2020channel}.
Given that the estimation process of the AOA at RIS for each RIS-reflected path is the same, we only take the $l$-th RIS as an example in this subsection.

According to the path gain model of the UE-RIS-AP paths given as \eqref{eq:path2-gain-sum-farfield}, the total path gain $g_{3,l}$ of the $l$-th RIS-reflected path has been estimated in Sec. \ref{sec-nomp}; 
according to Eq. \eqref{eq:path2-gain-sub-farfield}, the path gain of each RIS element in the far field is the same \cite{tang2022path};
the AOA at the AP is known.
Thus, only the AOA at the RIS is to be estimated.
However, the total path gain $g_{3,l}^{\circ}$ is the product of two complex number $g_{3,l}^{\rm{s}\circ}$ and $\mathbf{a}_{\rm{r}}^H(\varphi_{3,l}^{\rm{rp}\circ},\vartheta_{3,l}^{\rm{rp}\circ})\boldsymbol{\Omega}_l\mathbf{a}_{\rm{r}}(\varphi_{3,l}^{\rm{ru}\circ},\vartheta_{3,l}^{\rm{ru}\circ})$.
Thus, it is impossible to uniquely determine the AOA $(\varphi_{3,l}^{\rm{ru}\circ},\vartheta_{3,l}^{\rm{ru}\circ})$ at the RIS and $g_{3,l}^{\rm{s}\circ}$ from a single measurement of channel response.
According to \cite{deepak2020channel}, by sounding the channel with $B$ different phase shifts $\boldsymbol{\Omega}_{l,b}$, we obtain
\begin{equation}\label{eq:path2-gain-adjust}
g_{3,l,b}^\circ=g_{3,l}^{\rm{s}\circ}\mathbf{a}_{\rm{r}}^H(\varphi_{3,l}^{\rm{rp}\circ},\vartheta_{3,l}^{\rm{rp}\circ})\boldsymbol{\Omega}_{l,b}\mathbf{a}_{\rm{r}}(\varphi_{3,l}^{\rm{ru}\circ},\vartheta_{3,l}^{\rm{ru}\circ}),
\end{equation}
where $b = 1, 2, \ldots, B$ and $B\ge 2$.
By calculating $g_{3,l,1}^\circ / g_{3,l,b}^\circ$ with $b\ge 2$, we obtain 
\begin{equation}\label{eq:path2-gain-bizhi}
\begin{aligned}
\mathbf{a}_{\rm{r}}^H(\varphi_{3,l}^{\rm{rp}\circ},\vartheta_{3,l}^{\rm{rp}\circ})(g_{3,l,1}^\circ \boldsymbol{\Omega}_{l,b} & - g_{3,l,b}^\circ \boldsymbol{\Omega}_{l,1}) * \\
&\mathbf{a}_{\rm{r}}(\varphi_{3,l}^{\rm{ru}\circ},\vartheta_{3,l}^{\rm{ru}\circ}) = 0.
\end{aligned}
\end{equation}
With the existence of measurement noise in channel sounding and estimation error in the path gains, we have
\begin{equation}
\begin{aligned}
\mathbf{a}_{\rm{r}}^H(\varphi_{3,l}^{\rm{rp}\circ},\vartheta_{3,l}^{\rm{rp}\circ})(g_{3,l,1} \boldsymbol{\Omega}_{l,b} & - g_{3,l,b} \boldsymbol{\Omega}_{l,1}) * \\
&\mathbf{a}_{\rm{r}}(\varphi_{3,l}^{\rm{ru}\circ},\vartheta_{3,l}^{\rm{ru}\circ}) = e_b.
\end{aligned}
\end{equation}
Then, the problem of estimating $(\varphi_{3,l}^{\rm{ru}\circ},\vartheta_{3,l}^{\rm{ru}\circ})$ is transformed to the problem of solving the non-linear equations, given as
\begin{equation}\label{equation-111}
\mathbf{A}\mathbf{a}_{\mathrm{r}}\left(\varphi_{3,l}^{\rm{ru}\circ},\vartheta_{3,l}^{\rm{ru}\circ}\right)=\mathbf{e},
\end{equation}
where
\begin{equation}
\mathbf{A} = \left[\begin{array}{c}
\mathbf{a}_{\mathrm{r}}^{H}\left(\varphi_{3,l}^{\rm{rp}\circ},\vartheta_{3,l}^{\rm{rp}\circ}\right)\left(g_{3,l,1} \boldsymbol{\Omega}_{l,2}-g_{3,l,2} \boldsymbol{\Omega}_{l,1}\right) \\
\mathbf{a}_{\mathrm{r}}^{H}\left(\varphi_{3,l}^{\rm{rp}\circ},\vartheta_{3,l}^{\rm{rp}\circ}\right)\left(g_{3,l,1} \boldsymbol{\Omega}_{l,3}-g_{3,l,3} \boldsymbol{\Omega}_{l,1}\right) \\
\vdots \\
\mathbf{a}_{\mathrm{r}}^{H}\left(\varphi_{3,l}^{\rm{rp}\circ},\vartheta_{3,l}^{\rm{rp}\circ}\right)\left(g_{3,l,1} \boldsymbol{\Omega}_{l,B}-g_{3,l,B} \boldsymbol{\Omega}_{l,1}\right)
\end{array}\right],
\end{equation}
$\mathbf{e}$ denotes the error vector. 
Notably, the matrix $\mathbf{A}\in\mathbb{C}^{(B-1)\times M_{\rm{r}}}$ is typically full row rank, since the phase shifts of RIS elements are randomly chosen.
By employing the singular value decomposition (SVD) of $\mathbf{A}$ and considering the left singular vectors that corresponds to the $(B-1)$ largest singular values of $\mathbf{A}$, we can obtain the basis of $\mathcal{R}(\mathbf{A}^H)$, given as $\mathbf{U} \in \mathbb{C}^{M_{\rm{r}} \times (B-1)}$, where $\mathcal{R}(\mathbf{A}^H)$ denotes the column span of $\mathbf{A}^H$.
In the noiseless conditions where $\mathbf{e} = \mathbf{0}$, we have $\|\mathbf{U}^H\mathbf{a}_{\rm{r}}\left(\varphi_{3,l}^{\rm{ru}\circ},\vartheta_{3,l}^{\rm{ru}\circ}\right)\|^2=0$.
Therefore, with the existence of noise, we can obtain the estimates of the AOA at the RIS by computing the locations of the peaks of the pseudo spectrum
\begin{equation}\label{333}
\mathbf{P}\left(\varphi,\vartheta\right) = \| \mathbf{U}^H \mathbf{a}_{\rm{r}}\left(\varphi,\vartheta\right) \|^{-2}.
\end{equation}
In practice, the pseudo spectrum given in \eqref{333} is obtained by sweeping over $\left(\varphi,\vartheta\right)$ with plenty of sampling points, and the number of sampling points per $\pi/2$ is denoted as $N$, which influences the estimation accuracy of the AOAs at the RISs.

\subsection{UE Localization}
\label{sec-ls}
In this subsection, the AP and RISs are treated as the reference points together, and their locations are denoted as $\mathbf{r}_l=[x^{\rm{r}}_l,y^{\rm{r}}_l,z^{\rm{r}}_l]^T$, where $l = 1, 2, \ldots, L$ and $L\le L_3+1$.
Given that the locations of the reference points are known, and the AOAs at them have been estimated in Sec. \ref{sec-nomp} and \ref{sec-pseudo}, the LS estimator can locate the UE.
The estimated AOA at the $l$-th reference point is denoted as $(\phi_l, \theta_l)$, whose real value is given as $(\phi_l^\circ, \theta_l^\circ)$.
Here, we assume that $\phi_l = \phi_l^\circ + \Delta \phi_l$, and $\theta_l = \theta_l^\circ + \Delta \theta_l$, where $\Delta \phi_l$ and $\Delta \theta_l$ are both additive zero mean Gaussian noise with the variance $\sigma^2$.
The angles $(\phi_l^\circ, \theta_l^\circ)$ specify an orthonormal basis in the 3-D space, given as
\begin{equation}
\begin{aligned}
&\mathbf{d}(\phi_l^\circ\!, \theta_l^\circ)&\!\!\!\!\!=&[\cos (\phi_l^\circ)\! \sin (\theta_l^\circ)\!, \sin (\phi_l^\circ)\! \sin (\theta_l^\circ)\!, \cos (\theta_l^\circ)]^{T}\!, \\
&\mathbf{c}(\phi_l^\circ\!, \theta_l^\circ)&\!\!\!\!\!=&[-\sin (\phi_l^\circ), \cos (\phi_l^\circ), 0]^{T}, \\
&\mathbf{v}(\phi_l^\circ\!, \theta_l^\circ)&\!\!\!\!\!=&[-\cos (\phi_l^\circ) \cos (\theta_l^\circ), -\sin (\phi_l^\circ) \cos (\theta_l^\circ), \\
&&& \quad\quad\quad\quad\quad\quad\quad\quad\quad\quad\quad\quad \sin (\theta_l^\circ)]^{T},
\end{aligned}
\end{equation}
where vector $\mathbf{d}(\phi_l^\circ, \theta_l^\circ)$ is the unit-norm vector that points from the UE to the $l$-th reference point.
Then, we can have 
\begin{equation}
\begin{aligned}
&\mathbf{c}^T(\phi_l^\circ, \theta_l^\circ)\mathbf{r}_l = \mathbf{c}^T(\phi_l^\circ, \theta_l^\circ)\mathbf{u}^\circ,\\
&\mathbf{v}^T(\phi_l^\circ, \theta_l^\circ)\mathbf{r}_l = \mathbf{v}^T(\phi_l^\circ, \theta_l^\circ)\mathbf{u}^\circ,
\end{aligned}
\end{equation}
by utilizing the orthogonality of the basis vectors.
For all the $L$ reference points, we have 
\begin{equation}\label{25}
\underbrace{\left[\begin{array}{c}
\mathbf{c}^T(\phi_1^\circ, \theta_1^\circ)\mathbf{r}_1\\
\mathbf{v}^T(\phi_1^\circ, \theta_1^\circ)\mathbf{r}_1\\
\vdots \\
\mathbf{c}^T(\phi_L^\circ, \theta_L^\circ)\mathbf{r}_L\\
\mathbf{v}^T(\phi_L^\circ, \theta_L^\circ)\mathbf{r}_L
\end{array}\right]}_{\mathbf{y}^\circ \in \mathbb{R}^{2L\times 1}} = 
\underbrace{\left[\begin{array}{c}
\mathbf{c}^T(\phi_1^\circ, \theta_1^\circ)\\
\mathbf{v}^T(\phi_1^\circ, \theta_1^\circ)\\
\vdots \\
\mathbf{c}^T(\phi_L^\circ, \theta_L^\circ)\\
\mathbf{v}^T(\phi_L^\circ, \theta_L^\circ)
\end{array}\right]}_{\mathbf{G}^\circ \in \mathbb{R}^{2L\times 3}} \mathbf{u}^\circ.
\end{equation}
By replacing the noise-free parameters $(\phi_l^\circ, \theta_l^\circ)$ in $\mathbf{y}^\circ$ and $\mathbf{G}^\circ$ with the estimated parameters $(\phi_l, \theta_l)$, we can define the error vector
\begin{equation}\label{error-vector}
\mathbf{e}_0 = \mathbf{y} - \mathbf{Gu}^\circ,
\end{equation}
where $\mathbf{y}$ and $\mathbf{G}$ are the noisy counterparts of $\mathbf{y}^\circ$ and $\mathbf{G}^\circ$, respectively.
Then, the LS solution of $\mathbf{u}^\circ$ can be obtained by \cite{kay1993fundamentals}
\begin{equation}\label{ls}
\mathbf{u} = \left(\mathbf{G}^{T} \mathbf{G}\right)^{-1} \mathbf{G}^{T} \mathbf{y}.
\end{equation}
Notably, the LS estimator requires at least three equations in \eqref{25} to estimate $\mathbf{u}^\circ$, or equivalently, at least two reference points.
Finally, the proposed channel estimation and localization schemes are summarized in Algorithm \ref{alg1}.

\begin{algorithm}   
    \caption{\textbf{: Pseudocode of the Proposed Channel Estimation and Localization Schemes}}\label{alg1}  
    \begin{algorithmic}[1]  
        \Require $L_{\rm{max}}$, $E_0$, $\mathbf{r}_l$, $\boldsymbol{\Omega}_{l,b}$ and $\mathbf{h}_b$, where $l = 1,2,\ldots,L$ and $b=1,2,\ldots,B$.
        \Ensure  $\mathbf{u}$.
        \For {$b=1$ to $B$}
        \State $\mathcal{P}_{b,0} = \varnothing$, $l = 0$.
        \For  {$k=1$ to $L_{\rm{max}}$}
        \State Calculate\ \ the\ \ residual\ \ channel\ \ response 
        \Statex \quad \quad \quad \quad $\mathbf{h}_{b,k}^{\rm{r}}$ by \eqref{residual}.
        \If {$|\mathbf{h}_{b,k}^{\rm{r}}| < E_0$}
        \State Break.
        \EndIf 
        \State Employ the NOMP algorithm to estimate 
        \Statex \quad \quad \quad \quad the\ \  parameters\ \  of\ \  the\ \  $k$-th\ \  path\ \  $(g_{b,k}$, 
        \Statex \quad \quad \quad \quad $\omega_{b,k}$, $\psi_{b,k})$, and renew the set of param-
        \Statex \quad \quad \quad \quad eters as $\mathcal{P}_{b,k}$.
        \EndFor
        \State{\bf end}
        \State Match the estimates in $\mathcal{P}_{b,K_0}$ with the known 
        \Statex \quad \quad \quad AOA at the AP of the $l$-th RIS path $(\omega_{3,l}^\circ$, 
        \Statex \quad \quad \quad $\psi_{3,l}^\circ)$, and obtain $g_{3,l,b}$ of the $l$-th RIS path
        \Statex \quad \quad \quad for the $b$-th channel sounding.
        \State Calculate the AOA at the AP of the LOS path 
        \Statex \quad \quad \quad $\left(\phi_{1,1,b}, \theta_{1,1,b}\right)$ for the $b$-th channel sound- 
        \Statex \quad \quad \quad ing by \eqref{222}.
        \EndFor
        \State{\bf end}
        \State Average the estimated AOAs at the AP $\left(\phi_{1,1,b}, \theta_{1,1,b}\right)$ of different channel soundings and obtain $\left(\phi_{1,1}, \theta_{1,1}\right)$ by \eqref{average}.
        \For  {$l=1$ to $L$}
        \State Calculate the pseudo spectrum given in \eqref{333}
        \Statex \quad \quad \quad by using the phase shifts $\boldsymbol{\Omega}_{l,b}$ and the esti-
        \Statex \quad \quad \quad mated path gains $g_{3,l,b}$ of different channel 
        \Statex \quad \quad \quad soundings.
        \State Find the peak of the pseudo spectrum and ob-
        \Statex \quad \quad \quad tain the estimate of the AOA at the $l$-th RIS
        \Statex \quad \quad \quad $(\phi_{3,l}^{\rm{ru}}, \theta_{3,l}^{\rm{ru}})$.
        \EndFor
        \State{\bf end}
        \State Determine the location of UE with $\mathbf{r}_l$ and the estimated AOAs $(\phi_{3,l}^{\rm{u}}, \theta_{3,l}^{\rm{u}})$ at different reference points by employing the LS estimator, given as \eqref{ls}.
    \end{algorithmic}
\end{algorithm}

{\remark{Compared with the systems without RISs, the proposed system with multiple RISs can achieve localization of UE with a single AP.
The RISs can intelligently tune the radio environments and serve as additional reference points, while the hardware cost and energy consumption are relatively low.
Moreover, the proposed localization scheme can work even without LOS paths, and the communication capacity and coverage are believed to be enhanced by taking advantage of RISs.}}

\section{Numerical Results}
\label{sec:result}

In this section, we evaluate the performance of the proposed channel estimation and localization schemes in Sec. \ref{sec:localization}.
The size of the indoor environment is $10\times 10\times 3 \ \rm{m}^3$, as shown in Fig. \ref{fig:model}.
The working frequency $f = 5.24$ GHz.
The UPA of the AP employs $M_{\rm{p}} = M_{\rm{px}} \times M_{\rm{py}} = 10\times 10$ antenna elements.
The RIS is composed of $M_{\rm{r}} = M_{\rm{rx}} \times M_{\rm{ry}} = 10\times 10$ RIS elements.
The numbers of single and cyclic refinements in the NOMP algorithm are $R_{\rm{s}}=5$ and $R_{\rm{c}}=7$, respectively.
The energy threshold $E_0=-50\ \rm{dBm}$, and the estimation error threshold $T_0 = 0.1 \ \rm{rad}$.
The average RCS of scatterers is given as 0.6 $\text{m}^2$.
The AP is located at $\mathbf{p} = [2,2,3]^T\ \rm{m}$.
There are two RISs located at $[0,5,2]^T\ \rm{m}$ and $[5,0,2]^T\ \rm{m}$, respectively.
Notably, the distances between UE and RISs, as well as RISs and AP, may influence the performance of the proposed algorithms, since received signal power is negatively related to propagation distances.
Therefore, the UE location is randomly chosen from the region $\left\{[x, y, z]^{T}: 4 \leqslant x \leqslant 6,4 \leqslant y \leqslant 6, z=1\ \text{(Unit: m)}\right\}$ to release the effects of various distances.
The size of RIS element is typically of subwavelength scale within the range of $[\lambda/10, \lambda/2]$ \cite{tang2020wireless}, and we specify the RIS element size by $d_{\rm{x}}=d_{\rm{y}}=\lambda/2$ in this study.
The numbers of the paths of type two and type four are both randomly chosen from $\{3,4,5\}$.
The accuracies of localization and channel estimation are both evaluated by the root mean square error (RMSE), \textit{e.g.}, $\operatorname{RMSE}(\mathbf{u})=\sqrt{\sum_{t=1}^{T_{\rm{MC}}}\left\|\mathbf{u}_{t}-\mathbf{u}^{\circ}\right\|^{2} / T_{\rm{MC}}}$, where $T_{\rm{MC}}$ is the number of Monte Carlo simulations, and $\mathbf{u}_{t}$ is the estimate of $\mathbf{u}^\circ$ at the $t$-th simulation.
In the pseudo spectrum, the number of sampling points per $\pi/2$ is $N=200$.
We denote the transmit power as $P$ and assume that the additive noise power in the indoor environment is $-83 \ \rm{dBm}$ by considering RF chain impairments and low-bit quantization noise \cite{yang2020fast}.

\subsection{Pseudo Spectrum vs. Transmit Power}
\begin{figure}[t]
\centering
\captionsetup{font=footnotesize}
\begin{subfigure}[b]{0.49\linewidth}
\centering
\includegraphics[width=0.99\linewidth]{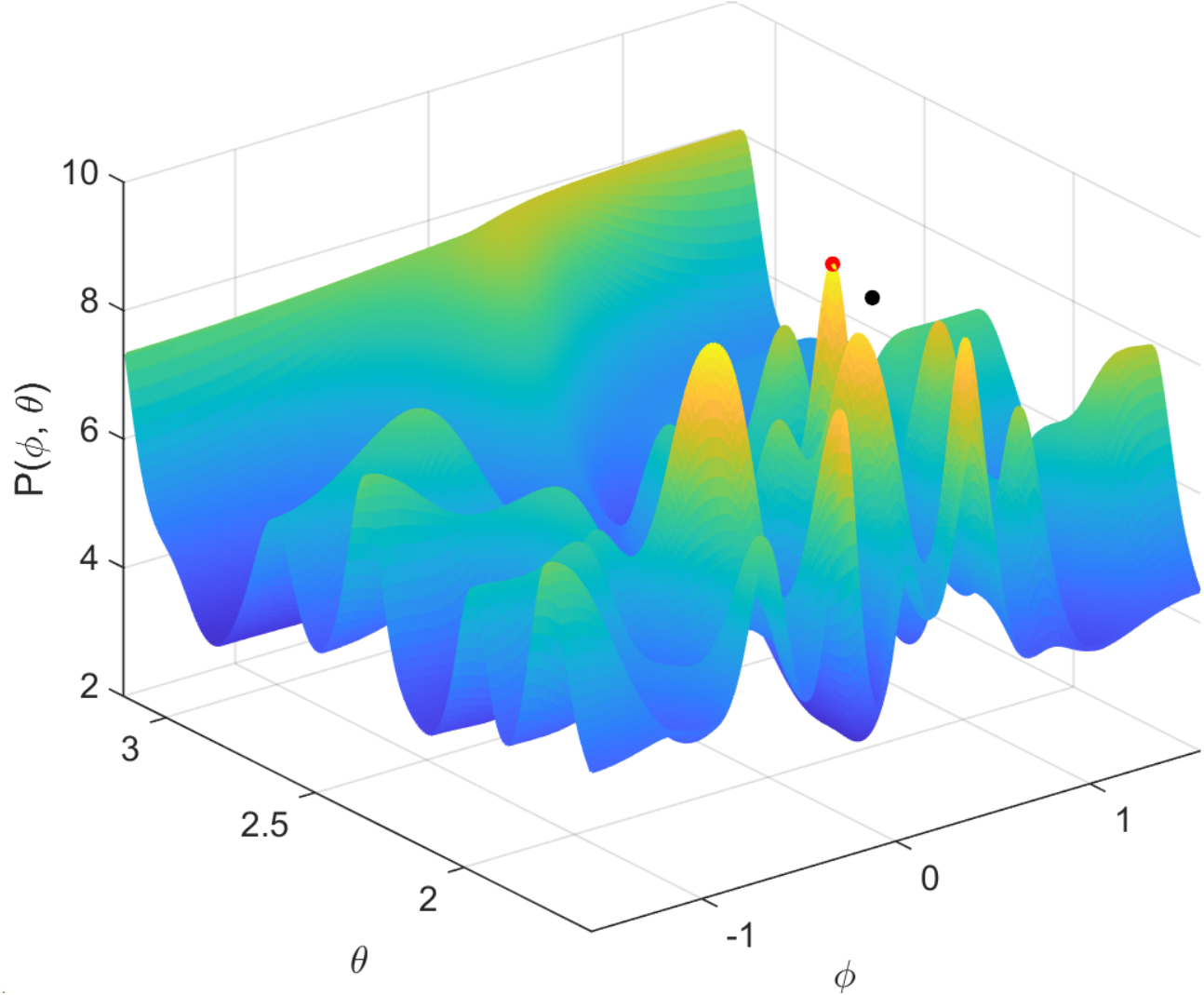}
\caption{$P=0\ \rm{dBm}$}
\vspace{0.2cm}
\label{fig1a}
\end{subfigure}
\begin{subfigure}[b]{0.49\linewidth}
\centering
\includegraphics[width=0.99\linewidth]{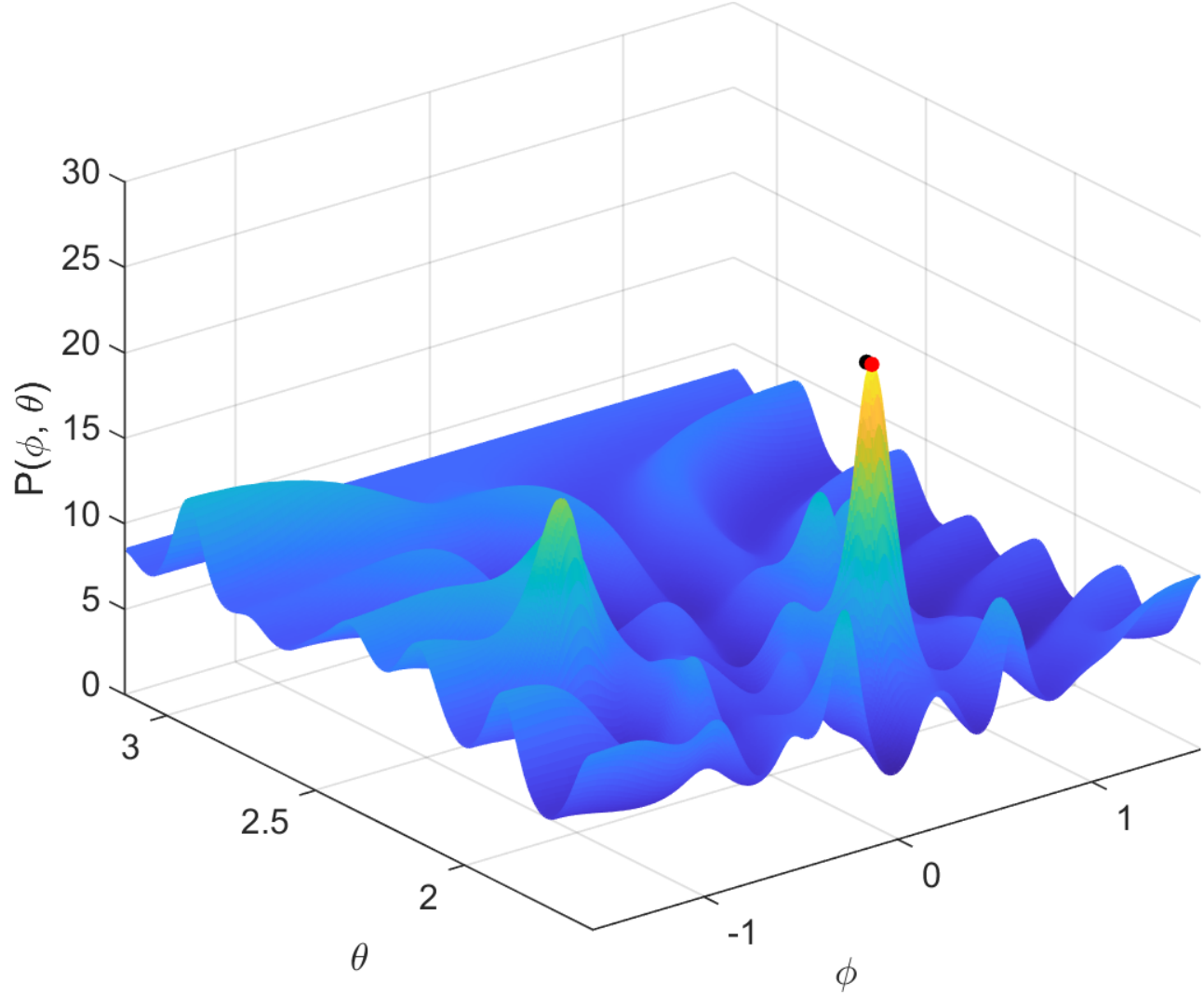}
\caption{$P=10\ \rm{dBm}$}
\vspace{0.2cm}
\label{fig1b}
\end{subfigure}
\begin{subfigure}[b]{0.49\linewidth}
\centering
\includegraphics[width=0.99\linewidth]{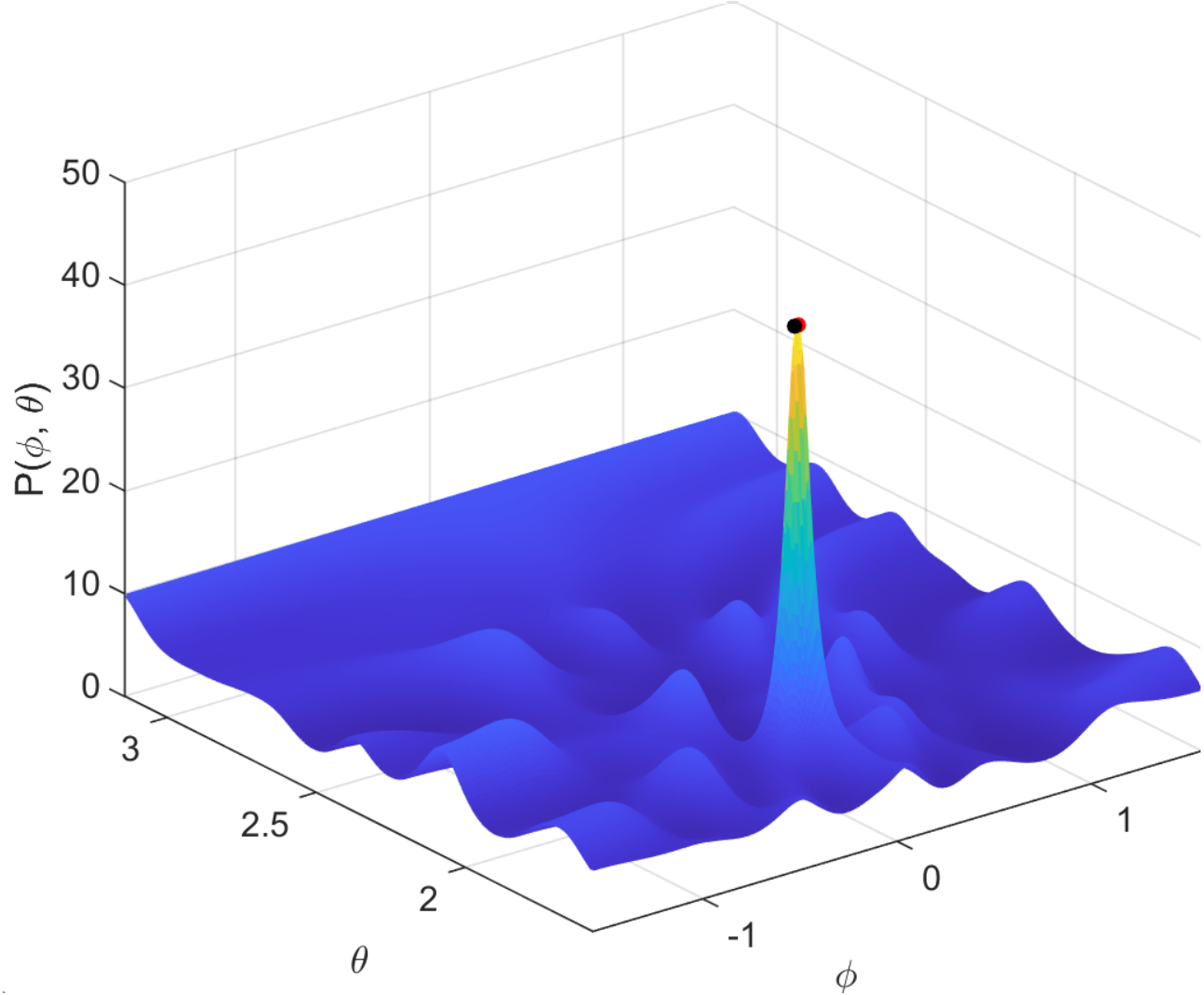}
\caption{$P=20\ \rm{dBm}$}
\label{fig1c}
\end{subfigure}
\begin{subfigure}[b]{0.49\linewidth}
\centering
\includegraphics[width=0.99\linewidth]{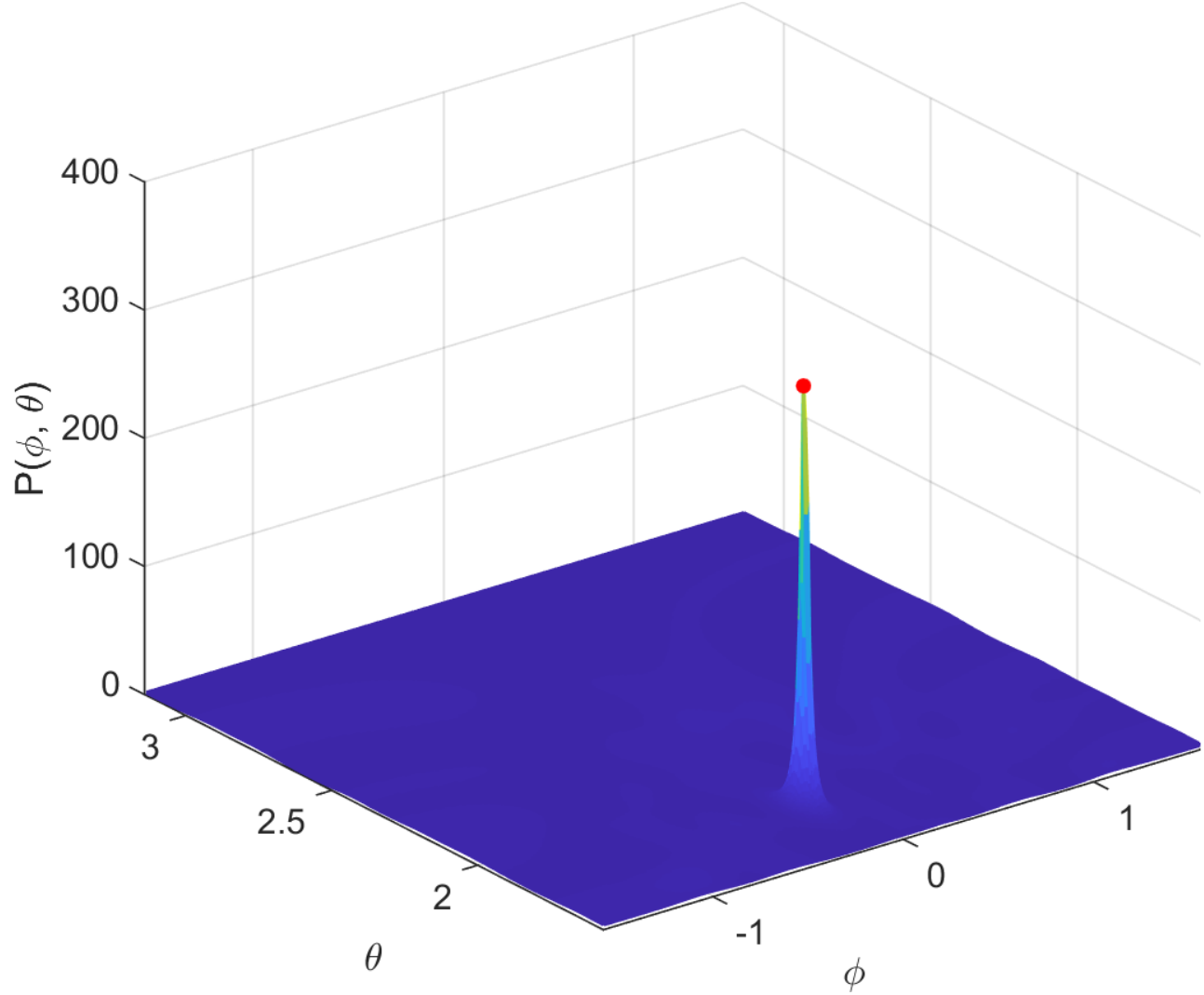}
\caption{$P=30\ \rm{dBm}$}
\label{fig1d}
\end{subfigure}
\caption{Pseudo spectrum with different transmit power $P$, where the units of $\phi$ and $\theta$ are in rad, the red point represents the peak value of the pseudo spectrum, and the black point represents the true location of AOA at RIS 1.}
    \label{fig1}
\end{figure}

In this subsection, we present the pseudo spectrum of RIS 1 given in \eqref{333} with different transmit power ($P=0, 10, 20, 30 \ \rm{dBm}$).
The results shown in Fig. \ref{fig1} illustrate how the transmit power influences the accuracy of channel estimation.
In Fig. \ref{fig1a} where the transmit power is relatively small, the channel response is submerged in the noise, and the pseudo spectrum goes up and down randomly, making the proposed channel estimation scheme ineffective.
In Fig. \ref{fig1b} and Fig. \ref{fig1c} where the transmit power increases gradually, the peak value of the pseudo spectrum results in a reasonable estimate of the AOA at RIS.
Moreover, the higher transmit power realizes a larger peak value of the pseudo spectrum and more accurate estimates of AOA.
In Fig. \ref{fig1d} where the transmit power is adequately large, the pseudo spectrum achieves an extremely large peak value at the position of the true AOA, whereas the values at other locations are nearly zero.
Therefore, the proposed method of channel estimation in RIS-aided communication systems performs well with large transmit power and becomes ineffective when the measurement noise is relatively large.

\begin{figure}[t]
\centering
\captionsetup{font=footnotesize}
\begin{subfigure}[b]{\linewidth}
\centering
\includegraphics[width=0.99\linewidth]{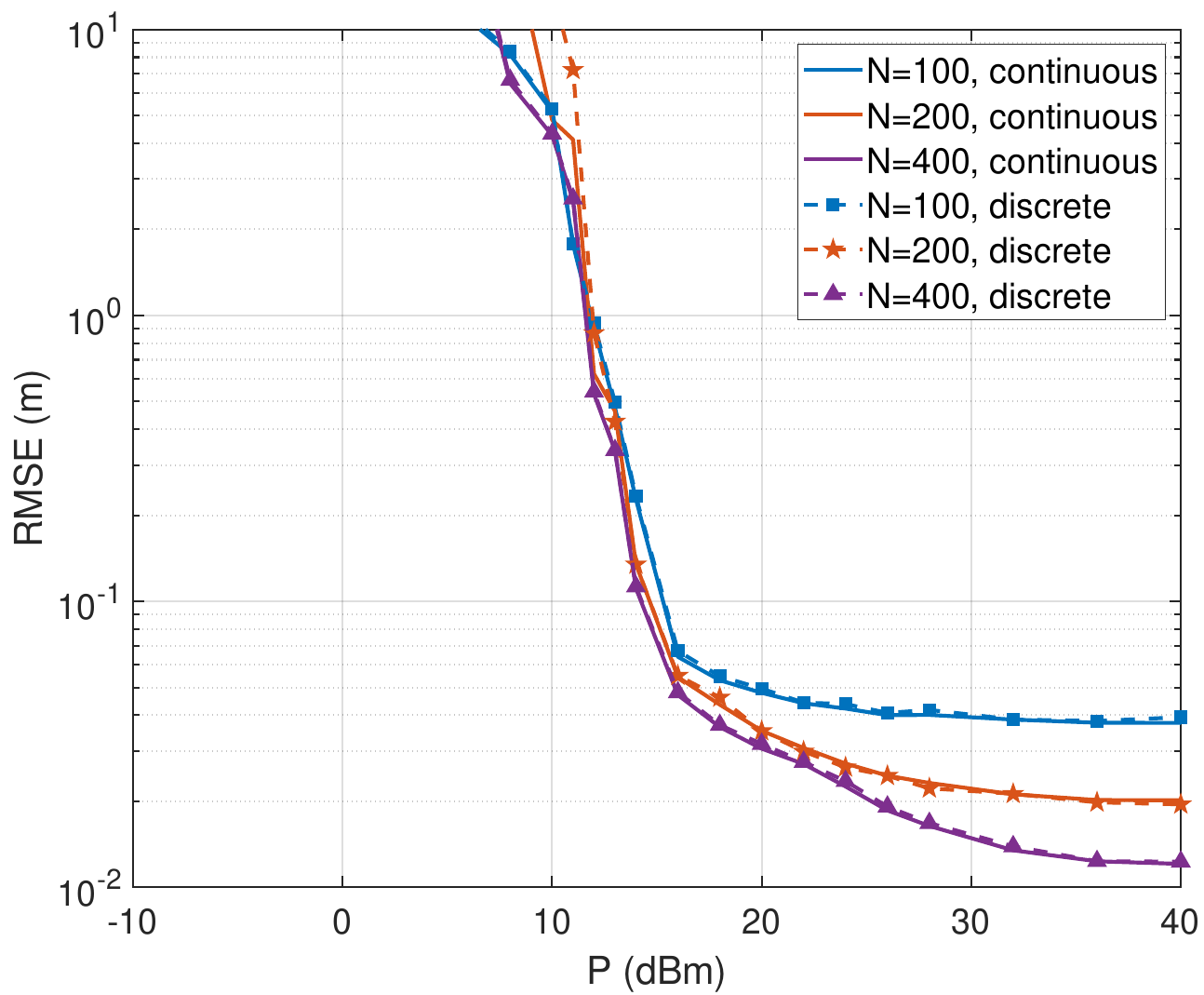}
\caption{UE position}
\label{fig2a}
\end{subfigure}
\begin{subfigure}[b]{\linewidth}
\centering
\includegraphics[width=0.99\linewidth]{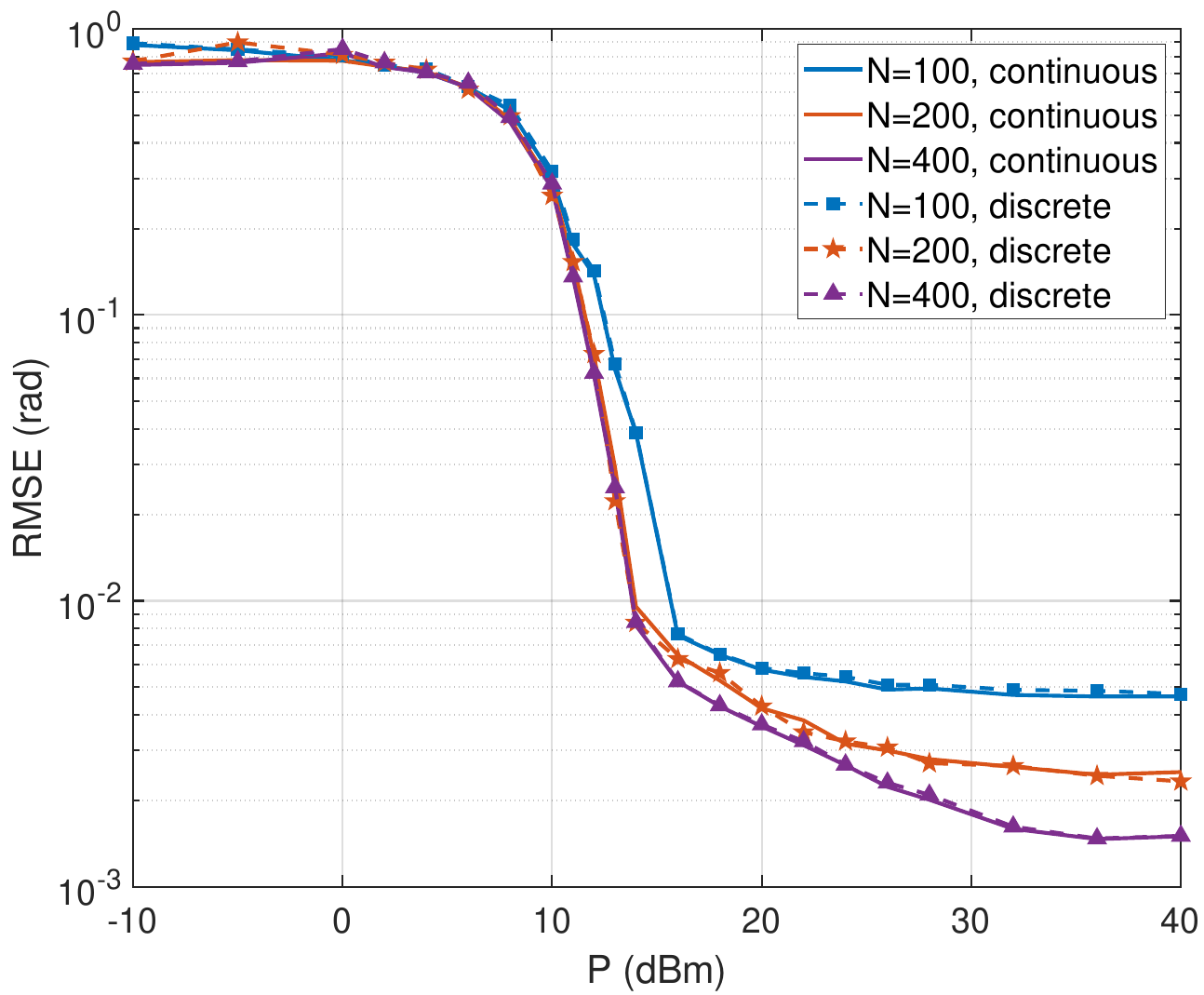}
\caption{Azimuth AOA at RIS1}
\label{fig2b}
\end{subfigure}
\caption{RMSEs of the locations and angles with different transmit power.}
    \label{fig2}
\end{figure}

\subsection{Estimation Accuracy vs. Transmit Power}\label{sec-result2}
In this subsection, we evaluate the performances of the proposed channel estimation and localization schemes with RIS-reflected paths.
Moreover, the performances between the continuous RIS phase shifts ("continuous" in Fig. \ref{fig2}) and 1-bit discrete RIS phase shifts ("discrete" in Fig. \ref{fig2}) are compared.
The results shown in Fig. \ref{fig2} come from 1000 independent Monte Carlo simulations with $B=30$.
The effects of transmit power $P$ and the number of sampling points $N$ in the pseudo spectrum are discussed.
It can be seen that the localization accuracy in Fig. \ref{fig2a} is strongly related to the estimation accuracy of the AOAs at the RISs, shown in Fig. \ref{fig2b}.
Notably, only the RMSE of the azimuth AOA at the first RIS is shown in this subsection and other RMSEs of estimated angles have similar values and tendencies.
When the transmit power is relatively small ($P<13$ dBm in Fig. \ref{fig2}), the proposed method in Sec. \ref{sec:localization} cannot determine the AOAs at the RISs and the location of UE effectively.
Accordingly, the RMSEs are dominated by the transmit power and may become very large when $P<8$ dBm.
With the transmit power increasing, the RMSEs of the AOA and UE position decrease gradually and approach the floors.
When the transmit power is adequately large, the floors of the RMSEs are mainly limited by the number of sampling points $N$.
The larger sampling points can ensure higher accuracy when $P>13$ dBm but lead to higher computation overhead.
Therefore, the proper number of sampling points can be determined by considering the transmit power, computation overhead, and localization accuracy requirements.
Additionally, the simulation results have shown that the RIS with 1-bit discrete phase shifts, which can realize $2^{M_{\rm{r}}}$ ($2^{M_{\rm{r}}}>>B$) different configurations, achieves nearly the same performance as the RIS with continuous phase shifts, since the proposed channel estimation scheme only requires that the matrix $\boldsymbol{\Omega}_{l,b}$ varies among the $B$ channel soundings.
Both the RISs with continuous and discrete phase shifts can obtain multiple measurements under different radio environments to realize channel estimation and localization.

\begin{figure}[t]
\centering
\captionsetup{font=footnotesize}
\begin{subfigure}[b]{\linewidth}
\centering
\includegraphics[width=0.97\linewidth]{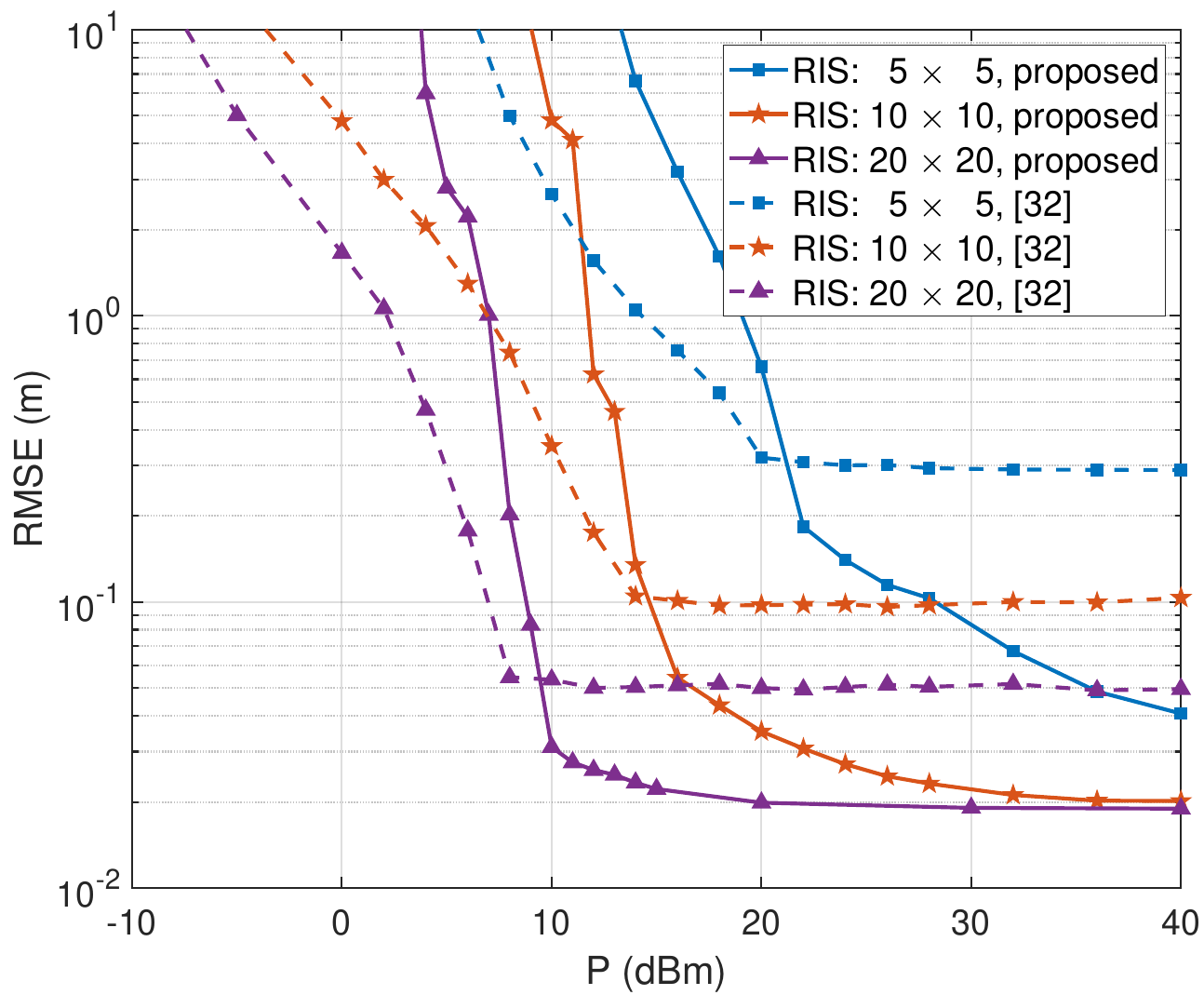}
\caption{UE position}
\vspace{0.0cm}
\label{fignewa}
\end{subfigure}
\begin{subfigure}[b]{\linewidth}
\centering
\includegraphics[width=0.97\linewidth]{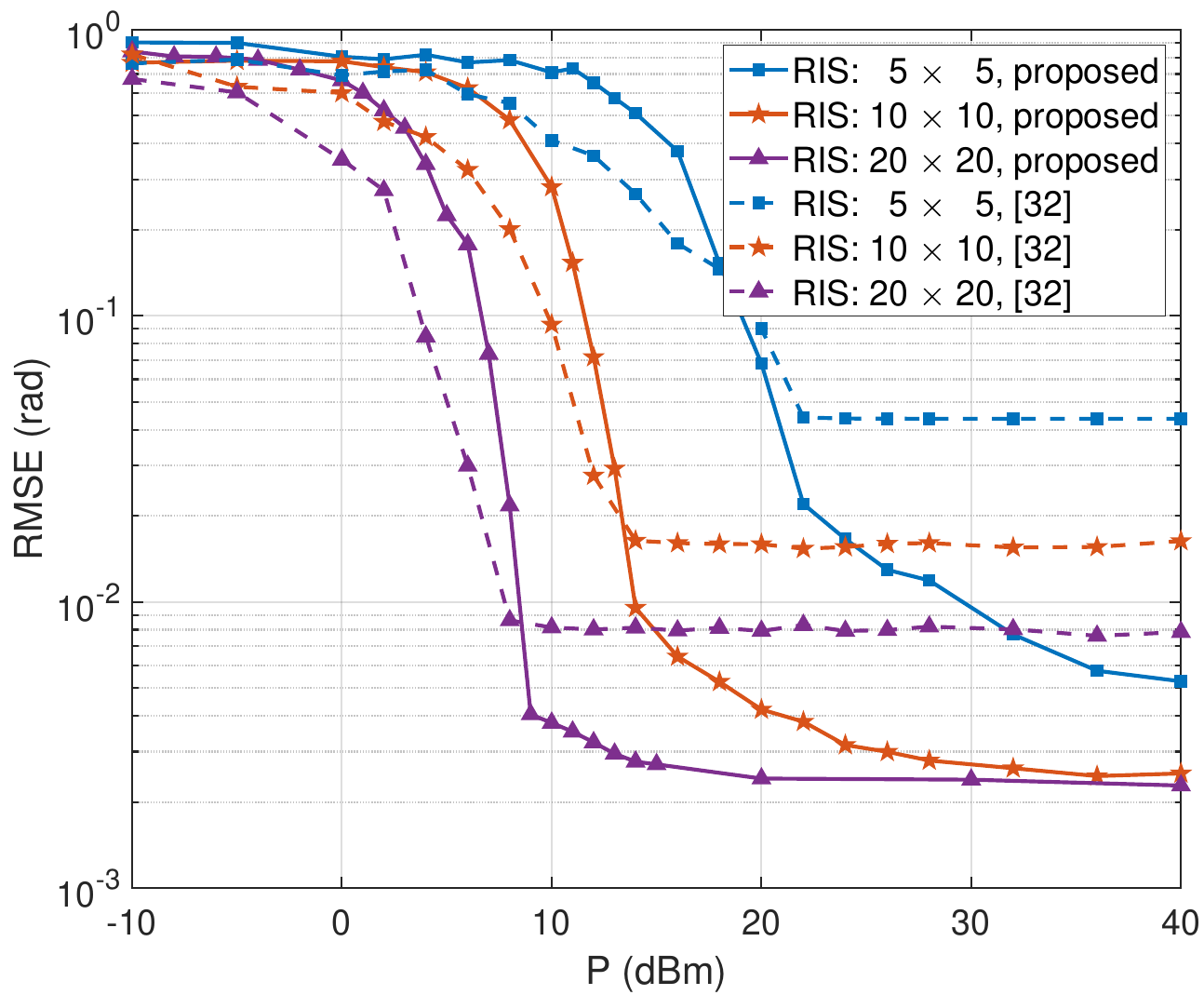}
\caption{Azimuth AOA at RIS1}
\label{fignewb}
\end{subfigure}
\caption{RMSEs of the locations and angles with different numbers of RIS elements, and the comparison with \cite{liu2021cascaded}.}
    \label{fig-new}
\end{figure}

\subsection{Estimation Accuracy vs. Number of RIS Elements}\label{sec-result-new}
In Sec. \ref{sec-result2}, large transmit power is demonstrated to achieve high localization performance.
However, large transmit power is not preferred in practical communication systems to reduce energy consumption \cite{gandotra2017green}.
In this subsection, we explore the effects of the number of RIS elements on the required transmit power.
The results shown in Fig. \ref{fig-new} come from 1000 independent Monte Carlo simulations with $B=30$, and the performances of the RISs with $5\times5$, $10\times10$, and $20\times20$ elements are compared.
The number of sampling points per $\pi/2$ in the pseudo spectrum is $N=200$.
It can be found that the transmit power required to achieve a certain estimation accuracy is degraded gradually with the number of RIS elements increasing.
A larger number of reflecting elements of RIS means larger dimensions of the steering vectors of RIS pointing to UE and AP in Eq. \eqref{equation-111}.
Consequently, the number of columns of the matrix $\mathbf{A}$ is larger.
That is, more information will be involved to obtain the AOAs at RISs, which can achieve channel estimation and localization with higher accuracy.

Additionally, we compare the estimation accuracy of the proposed channel estimation scheme with the method in \cite{liu2021cascaded}.
In contrast to our proposed scheme that first determines the path gains of RIS-reflected paths, and the estimation error may be propagated, the method in \cite{liu2021cascaded} avoids additional estimation error in the path gains, since the received signal vector was directly used for matching pursuit with the vector dimension being the number of channel soundings $B$.
Thus, the method in \cite{liu2021cascaded} can obtain higher accuracy than our proposed scheme with low transmit power, as illustrated by Fig. \ref{fig-new}.
However, with the increase of transmit power, the path gains of RIS-reflected paths can be estimated with high accuracy.
Moreover, in our proposed scheme, the dimension of the vectors in matching pursuit is $M_{\rm{r}}$, which is typically much larger than that of the method in \cite{liu2021cascaded}.
Therefore, the proposed channel estimation scheme outperforms the method in \cite{liu2021cascaded} with large transmit power.

\subsection{Estimation Accuracy vs. Number of Channel Soundings}
\begin{figure}[t]
\centering
\captionsetup{font=footnotesize}
\begin{subfigure}[b]{\linewidth}
\centering
\includegraphics[width=0.99\linewidth]{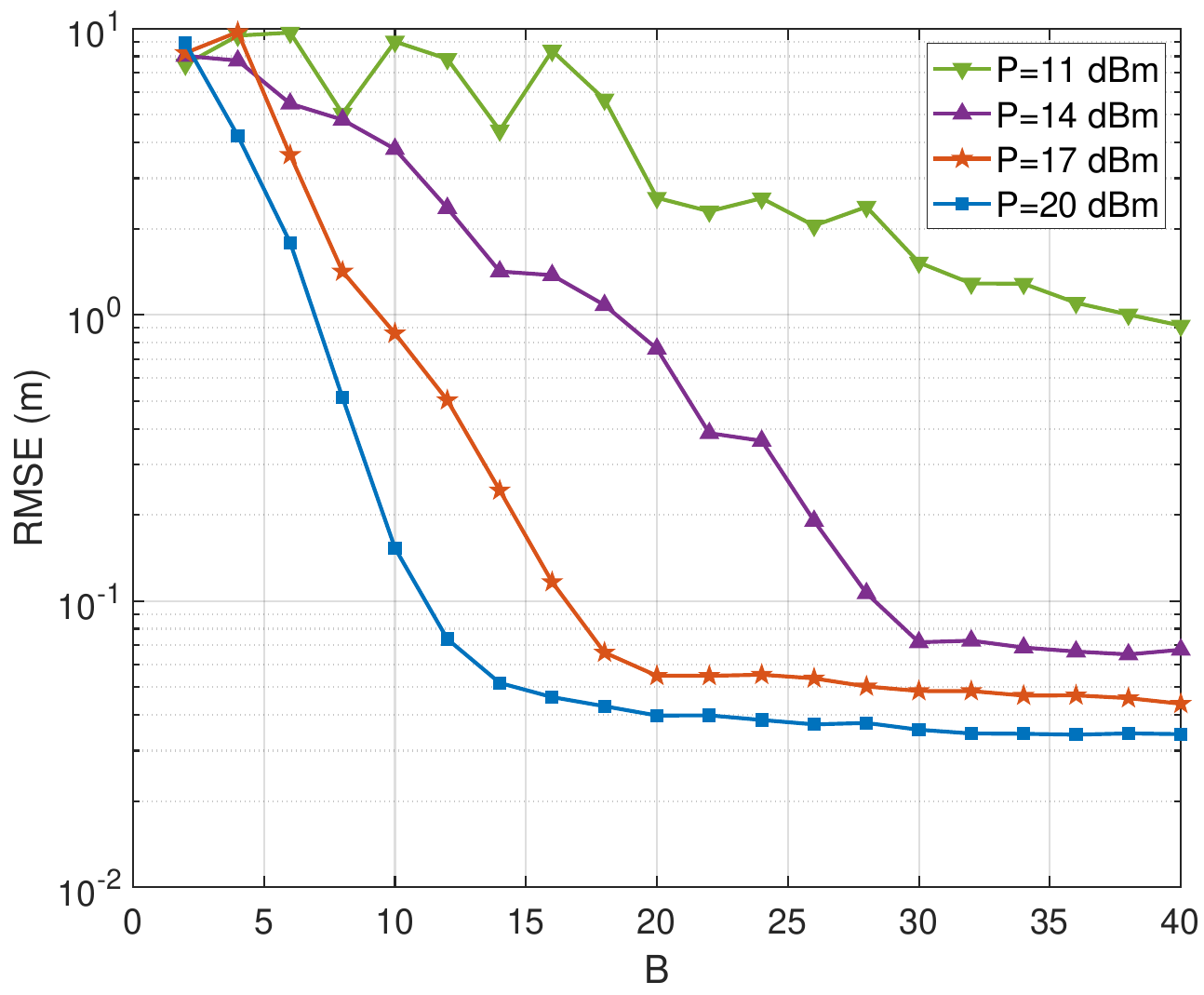}
\caption{UE position}
\label{fig3a}
\end{subfigure}
\begin{subfigure}[b]{\linewidth}
\centering
\includegraphics[width=0.99\linewidth]{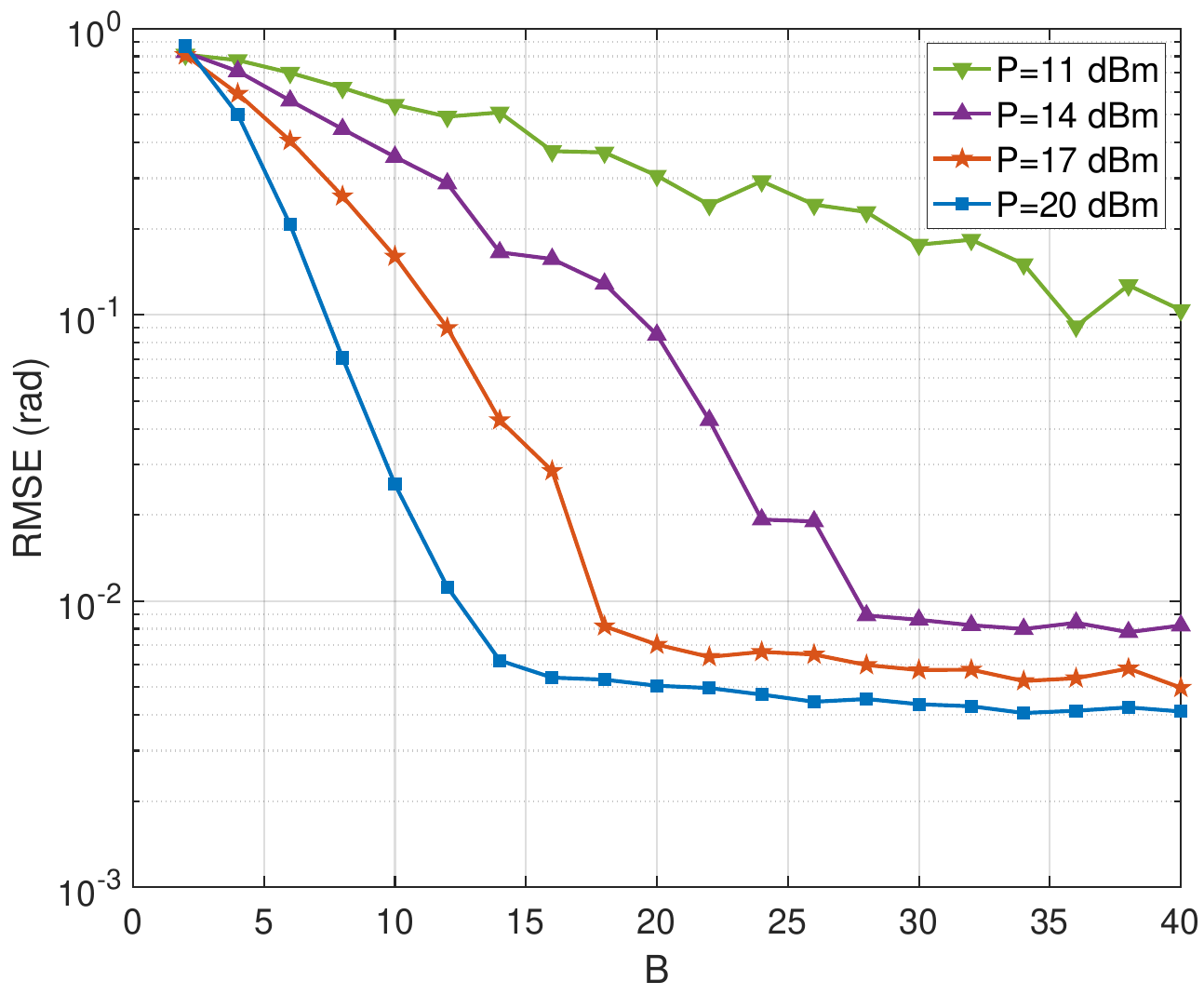}
\caption{Azimuth AOA at RIS1}
\label{fig3b}
\end{subfigure}
\caption{RMSEs of the locations and angles with different numbers of channel soundings.}
    \label{fig3}
\end{figure}

In the proposed algorithm in Sec. \ref{sec:localization}, the number of channel soundings $B$ is a key factor that influences the efficiency of localization.
A small number of channel soundings may be not adequate to obtain the AOAs at RISs with high accuracy, while a large value of $B$ may lead to the waste of time and computation resources.
Therefore, we discuss the influence of the number of channel soundings $B$ on the estimation accuracy in this subsection.
The LOS path is assumed to be obstructed and two RIS-reflected paths are ensured.
We set $N=200$, $T_{\rm{MC}}=1000$, and $P=11, 14, 17, 20$ dBm to obtain the simulation results shown in Fig. \ref{fig3}.
Similarly to the discussion in Sec. \ref{sec-result2}, a clear dependence of the localization accuracy on the channel estimation accuracy can be found by comparing Fig. \ref{fig3a} and Fig. \ref{fig3b}.
With the number of channel soundings increasing, the RMSEs of UE location and AOA decrease gradually and reach the floors.
The RMSEs with large transmit power can approach the floors with a small number of channel soundings.
Moreover, the larger the transmit power is, the lower the floors of RMSEs are.
Additionally, when the transmit power is extremely small, \textit{e.g.}, $P=11$ dBm, even 40 channel soundings of different phase shifts cannot reach the floors of the RMSEs.
Therefore, reaching the floors of the RMSEs requires a tradeoff between the transmit power and the number of channel soundings.

\subsection{Localization with Different Numbers of Reference points}

\begin{figure}
    \centering
    \includegraphics[width=0.99\linewidth]{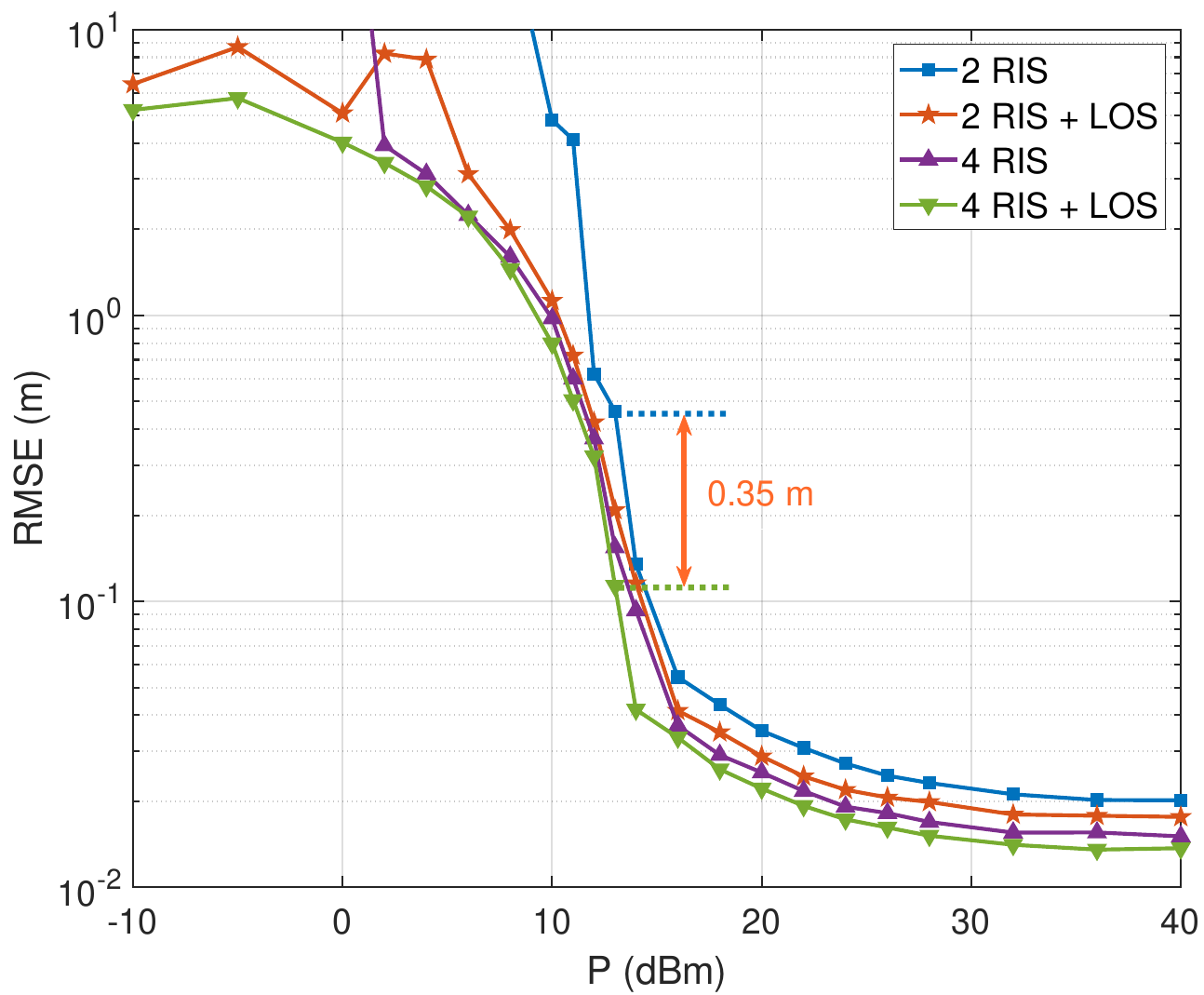} 
    \captionsetup{font=footnotesize}
    \caption{RMSEs of the locations and angles with different numbers of reference points.}
    \label{fig4}
\end{figure}

In this subsection, we discuss the influence of the number of reference points (\textit{i.e.}, the number of LOS and RIS-reflected paths) on localization accuracy.
The RMSEs of UE location with 2 RISs, 2 RISs plus a LOS path, 4 RISs, and 4 RISs plus a LOS path are shown in Fig, \ref{fig4}, where the locations of the third and fourth RISs are $[0,3,2]^T$ m and $[3,0,2]^T$ m, respectively.
The simulation results show that the existence of the LOS path and additional RIS-reflected paths can improve the location accuracy of UE with nearly all the transmit power.
Specifically, the location accuracy with 4 RISs plus a LOS path achieves the best performance in this subsection.
When the transmit power is very small ($P \le 10$ dBm with 2 RISs, and $P \le 0$ dBm with 4 RISs), the RMSE performances with no LOS paths are extremely bad.
However, the existence of LOS paths greatly degrades the RMSEs with small transmit power.
Moreover, the RMSE with 4 RISs plus a LOS path is much lower than that with 2 RISs when $10\le P\le 16$ dBm.
With the RMSEs gradually reaching the floors when $P\ge 16$ dBm, the performance gaps between the considered four scenarios become small.
In conclusion, the location accuracy of UE can be improved with the assistance of the LOS path and additional RIS-reflected paths, and the performance improvements with small transmit power are remarkable.

\section{Conclusion}
\label{sec:conclusion}

This study investigated the problem of RIS-aided localization with a single AP.
A two-stage channel estimation scheme that requires the RISs to tune the phase shifts to obtain multiple channel soundings was proposed to determine the AOAs at RISs.
For each channel sounding, the first stage employed the traditional NOMP algorithm to extract the multiple paths, and the LOS path and RIS paths can be identified.
Then, the second stage determined the AOAs at the RISs from the pseudo spectrum by using the estimated path gains of RIS-reflected paths.
Consequently, the AOAs at AP and RISs were utilized to estimate the location of UE with the LS estimator.
Simulation results showed that the proposed algorithm can achieve centimeter-level localization accuracy with large transmit power, and the localization accuracy was partly limited by the number of sampling points in the pseudo spectrum.
Moreover, the smaller transmit power required a larger number of channel soundings to reach the floors of RMSEs.
The proposed algorithms can be extended to scenarios when the UE is equipped with multiple antennas in future study, whereas the angle of departure at UE can be derived to help improve UE location accuracy, estimate antenna array orientation, and further determine scatterer locations.

\section*{NOTES}
${}^1$The far-field assumption requires that the distance between the UE and RISs should be larger than $d_0=2D^2/\lambda$, where $D$ is the largest dimension of the RISs. In the studied scenario in this paper, we have $d_0=2.86\ \rm{m}$. Additionally, the nearest distance between the UE moving region and RISs is $4\ \rm{m}$, and the distance between the AP and RISs is $3.75\ \rm{m}$. Thus, the UE and AP are both in the far-field region of RISs.

\hspace{-1em}${}^2$
The multiple-reflected paths, including UE-scatterer-RIS-AP paths, UE-RIS-scatterer-AP paths, UE-RIS1-RIS2-AP paths, and other paths that experience more reflections, result in neglectable signal energy compared with the single-reflected paths.
Moreover, the modelling of these multiple-reflected paths is tedious.
Therefore, we take the UE-scatterer-RIS-AP paths as an example to represent the influence of the multiple-reflected signals, whereas other types of multiple-reflected paths are considered additive noise in this system.

\footnotesize
\bibliographystyle{IEEEtran}
\bibliography{myref}

\end{document}